\documentclass{article}
\usepackage{amsfonts}
\usepackage[page]{appendix}
\usepackage{amsthm}
\usepackage{amsmath}
\usepackage{color}
\usepackage{xcolor}
\usepackage{dsfont}
\usepackage{graphicx}
\DeclareMathOperator*{\esssup}{ess\,sup}
\DeclareMathOperator*{\essinf}{ess\,inf}
\DeclareMathOperator*{\argmin}{arg\,min}

\theoremstyle{definition}
\newtheorem{theorem}{Theorem}[section]
\newtheorem{corollary}[theorem]{Corollary}
\newtheorem{lemma}[theorem]{Lemma}
\newtheorem{proposition}[theorem]{Proposition}

\newtheorem{definition}{Definition}[section]

\newtheorem{remark}{Remark}[section]
\newtheorem{example}{Example}[section]
\begin{document}

\sloppy
\allowdisplaybreaks

\title{Robust forward investment and consumption under drift and volatility uncertainties: A randomization approach\footnote{We express our gratitude to the two referees and the Associate Editor for their insightful and constructive comments.}}
\author{Wing Fung Chong\thanks{%
Department of Statistics and Actuarial Science, School of Computing and Data Science, The University of Hong Kong, Pokfulam, Hong Kong, China; email:
\texttt{chongwf@hku.hk}} \and Gechun Liang\thanks{%
Department of Statistics, The University of Warwick, Coventry CV4
7AL, U.K.; email: \texttt{g.liang@warwick.ac.uk} }}
\date{}
\maketitle

\begin{abstract}
This paper studies robust forward investment and consumption preferences and optimal strategies for a risk-averse and ambiguity-averse agent in an incomplete financial market with drift and volatility uncertainties. We focus on non-zero volatility and constant relative risk aversion forward preferences. Given the non-convexity of the Hamiltonian with respect to uncertain volatilities, we first construct robust randomized forward preferences through endogenous randomization in an auxiliary market. {Therein, w}e derive the corresponding optimal and robust investment and consumption strategies. Furthermore, we show that such forward preferences and strategies, developed in the auxiliary market, remain optimal and robust in the physical market, offering a comprehensive {analysis} for forward investment and consumption under model uncertainty.
\end{abstract}

{\section{Introduction}

In modeling preferences for optimal investment problems, Musiela and Zariphopoulou introduced the concept of \textit{forward preferences} in their seminal works \cite{MZ0,MZ-Kurtz,MZ-Carmona,MZ1,MZ2,MZ3}. This framework addresses key limitations of traditional expected utility theory, where static utility functions are typically used to model future preferences. Unlike classical approaches, forward preferences begin with an initial datum representing the agent's current preferences, which then evolve dynamically according to market conditions. This evolution is governed by two fundamental principles: the super-martingale property of sub-optimal strategies and the martingale property of optimal strategies. Crucially, this dynamic formulation ensures time-consistency of optimal strategies across all horizons, allowing investors to make optimal decisions without pre-specifying either the investment horizon or future preferences.

Over the past decades, substantial progress has been made in the development of forward preference theory. A central challenge has been the construction of such preferences. One prominent approach relies on stochastic partial differential equations (SPDEs), derived via a formal application of the It\^o-Wentzell formula. These SPDEs are fully nonlinear and degenerate, and their analysis remains technically challenging. Notable progress has been made using stochastic flow methods and convex duality, as in \cite{El_Karoui_2013,El_Karoui_2018}. For exponential-type forward preferences, convex duality offers more explicit solutions \cite{Zit}.
In the special case of zero-volatility (time-monotone) preferences, the SPDE reduces to a time-reversed ill-posed parabolic PDE. The solvability of this PDE hinges on Widder’s theorem, which characterizes positive solutions to parabolic equations; see \cite{Avanesyan_2018, Henderson_Hobson_2007,Nadtochiy_2017,Nadtochiy-Z, Shkolnikov_2016}.

For the construction of non-zero volatility preferences in general incomplete markets, a notable breakthrough in \cite{LZ} employed infinite-horizon ergodic backward stochastic differential equations (BSDEs) to construct homothetic forward preferences of exponential and power types. This approach has since been extended in various directions, including \cite{Broux-Quemerais_2024, Chong_2018, Hu2020,LLL}.

The forward theory has also been adapted to discrete-time models through predictable forward preferences \cite{Angoshtari2020}, with further developments in \cite{Angoshtari,Angoshtari_2024,Liang2023,Strub2021,Waldon_2024}. In parallel, forward preferences have been generalized to broader semimartingale settings \cite{Bo_2023,Choulli_2007}.
Applications of forward preferences span a wide range of areas, including:
equilibrium models \cite{El_Karoui_2021}, relative preferences \cite{Anthropelos2022,Zariphopoulou_2024}, insurance \cite{Chong_2018,Ng_Chong_2023}, dynamic risk measures \cite{Chong_2016,ZZ}, utility indifference pricing \cite{Anthropelos2014,Wang_2024}, optimal liquidation \cite{Wang_2024_b}, yield curve modeling \cite{El_Karoui_2022}, behavioral finance \cite{He_2021}, and pension policy \cite{ Anthropelos2025, Hillairet_2024}.

This paper makes two main contributions to the literature on the construction of forward preferences:

\noindent\textit{(i) Robust forward investment preferences under model uncertainty.}

We aim to construct {constant relative risk aversion} (CRRA) forward investment preferences of the form
\[
U(X_t^{\pi, c, b, \sigma}, t) = \frac{(X_t^{\pi, c, b, \sigma})^\kappa}{\kappa} e^{K_t},
\]
where \( X_t^{\pi, c, b, \sigma} \) denotes the wealth process at time \( t \) under trading strategy \( \pi \), consumption rate \( c \), drift \( b \), and volatility \( \sigma \). The parameter \( \kappa \in (0,1) \) captures the agent’s risk aversion, and \( K \) is an \textit{opportunity process} to be endogenously determined by an infinite-horizon BSDE. The driver of this BSDE is determined from the {saddle value of a Hamiltonian}.
In contrast to the model in \cite{LZ}, which constructs such preferences assuming the drift and volatility are known stochastic processes, our framework accounts for {model uncertainty}. Specifically, the agent is uncertain about the drift 
 and the \textit{idiosyncratic} volatility but is certain about the \textit{systematic} volatility, reflecting a practically motivated distinction: while aggregate market risks are typically observable and well-measured, stock-specific parameters are notoriously difficult to estimate with precision. 
We adopt a robust optimization approach, evaluating worst-case scenarios within defined uncertainty sets for drifts and idiosyncratic volatilities.

In the forward framework, the concept of robust forward investment preferences was initially developed by K{\"a}llblad, Ob{\l}{\'o}j, and Zariphopoulou in \cite{Kallblad_2018}. Their work focuses on model uncertainty represented by a set of equivalent probability measures, which corresponds to drift uncertainty in our setting via Girsanov's transformation (see also \cite{LLL, Lin_2020}). Saddle points in general exist under drift uncertainty. However, when volatility uncertainty is introduced, the resulting set of probability measures becomes mutually singular, leading to the non-existence of saddle points. This introduces substantial mathematical difficulties that require more sophisticated techniques. 

The most closely related work on forward preferences under volatility uncertainty is our previous paper \cite{Chong_2024}. However, the current paper departs fundamentally from \cite{Chong_2024}, both in its financial modeling and mathematical structure. In \cite{Chong_2024}, all market coefficients are uncertain, and the worst-case scenario corresponds to attainable saddle points that lie on the boundaries of the uncertainty sets. As a result, the worst-case market becomes a model with constant coefficients, which is essentially a complete market. This leads to a significant simplification: the associated BSDE reduces to an ODE, and the opportunity process $K$ becomes deterministic. Consequently, the forward preferences constructed in \cite{Chong_2024} are of the\textit{ zero-volatility} type.

In contrast to \cite{Chong_2024}, the market in our setting is generically incomplete, even after evaluating worst-case scenarios. This market incompleteness arises from the presence of unhedgeable risks, which cannot be replicated by the wealth process, as well as from the existence of systematic volatility components, which are modeled as known stochastic processes. As a result, the worst-case coefficients or saddle points do not exist, and the opportunity process 
$K$ becomes stochastic, leading to the construction of forward preferences with \textit{non-zero volatility}. Accordingly, the optimal and robust trading strategies comprise not only the myopic component, as in the zero-volatility case, but also hedging components for both idiosyncratic and systematic risks.

The nonexistence of worst-case coefficients or saddle points complicates the derivation of optimal investment strategies in robust optimization. When a saddle point exists, one can first maximize the Hamiltonian over the trading strategy $\pi$ for any fixed volatility $\sigma$, leveraging the quadratic structure of the Hamiltonian with respect to $\pi$ to obtain a closed-form solution, followed by minimizing over $\sigma$ to derive explicit optimal strategies. Without a saddle point, however, identifying optimal strategies becomes unclear. To address this, we extend the randomization technique from \cite{Tevzadze_2013}, who applied randomization for optimal investment under static preferences. By introducing randomization, the agent can select the worst-case volatility through a probability distribution, which in turn convexifies the Hamiltonian in the BSDE driver, ensuring saddle points as probability distributions. Consequently, the BSDE becomes a randomized infinite-horizon BSDE, resolving volatility uncertainty through optimization over distributions rather than pointwise minimization.

This randomization has important financial implications. It is endogenously driven by unhedgeable risks present in the market. Since the underlying risky assets are insufficient to fully hedge all the risks, the agent partially mitigates the hedging error by using them to randomize the volatility, thereby selecting the worst-case distribution of volatility rather than a single pointwise realization. In effect, the agent is said to operate in an auxiliary financial market induced by this randomization. The existence of a saddle point for the randomized Hamiltonian is crucial, as it enables the characterization of the optimal and robust investment strategies as well as the worst-case volatility distributions in the auxiliary market. We subsequently demonstrate that the forward preferences and the corresponding strategies, constructed in this auxiliary setting, also serve as robust forward investment preferences, as well as optimal and robust strategies, for the agent operating in the original, physical financial market.

\noindent\textit{(ii) Forward consumption preferences.}

The second contribution of this paper concerns forward consumption preferences. We aim to construct CRRA forward consumption preferences of the form 
\[
U^C(C_t, t) = \frac{C_t^{\kappa}}{\kappa} \lambda_t,
\]
where $C$ is the consumption, and $\lambda$ is a stochastic process to be determined. The process $\lambda$ reflects the agent’s evolving beliefs about future consumption preferences. While the opportunity process $K$ is required to satisfy an infinite-horizon BSDE, our results show that no such requirement is imposed on $\lambda$, other than being uniformly bounded and satisfying a structural condition. Specifically, to guarantee the joint existence of forward investment and consumption preferences, it suffices that the forward consumption preference is dominated by the initial investment preference. A typical example is when $\lambda$ acts as a stochastic discount factor, capturing the agent’s declining marginal utility from consumption over time.

Forward consumption preferences were first introduced by \cite{Berrier_2008}, who developed a forward framework for dynamic optimization of consumption without prespecifying the time horizon or intertemporal preferences, using convex duality for characterization. Related works by \cite{El_Karoui_2018} and \cite{Kallblad_2016} employed SPDEs to characterize forward consumption preferences, while \cite{El_Karoui_2024} linked forward investment preferences in a defaultable setting to forward investment and consumption preferences.

This paper is structured as follows: Section 2 formulates the problem by establishing the physical financial market and defining robust forward preferences. Section 3 introduces randomization and develops robust randomized forward preferences within the auxiliary financial market. Section 4 constructs non-zero volatility robust randomized CRRA forward preferences and derives the associated optimal and robust investment and consumption strategies in the auxiliary market. Section 5 demonstrates that these preferences and strategies also represent non-zero volatility robust CRRA forward preferences and optimal strategies in the physical market. All proofs are provided in Section 6, and Section 7 concludes.}

\section{Problem Formulation}
\subsection{The Physical Market}
Consider a financial market starting from the current time $t=0$. Let 
\[
\mathbb{W}:=(W_t^1,\dots,{W_t^n,\bar{W}_t,B_t^{1}},\dots,B_t^{n}), \quad t\geq 0,
\]
be a $\left(2n+1\right)$-dimensional Brownian motion defined on a probability space $(\Omega ,\mathcal{F},\mathbb{P})$. The independent Brownian components $(W^1,\dots,W^n,\bar{W})$ are the driving noises which can be hedged using underlying stocks in the market, while the other independent Brownian components $(B^{1},\dots,B^{n})$ model the noises which cannot be fully hedged. Assume that, for each $i=1,\dots,n$, $W^i$ and $B^i$ are correlated with a constant correlation coefficient $\rho^i\in[-1,1]$, i.e., 
$
\langle W^i_{\cdot},B^i_{\cdot}\rangle_t=\rho^i t,\ t\geq 0,
$
while $W^j$, $\bar{W}$, and $B^i$ for $i\neq j$ are independent. Denote by $\mathbb{F}=\{\mathcal{F}_{t}\}_{t\geq 0}$ the natural filtration of $\mathbb{W}$ after augmentation.

The market consists of a risk-free bond with a constant interest rate \( r \in \mathbb{R}_+ \) and \( n \) risky stocks with price processes \( (S_t^1, \dots, S_t^n) \), \( t \geq 0 \). For each \( i = 1, \dots, n \), the stock price \( S_t^i \) evolves according to the diffusion process:
\begin{equation}\label{stock}
\frac{dS_t^i}{S_t^i} = b_t^i \, dt + \sigma_t^i \, dW_t^i + \bar{\sigma}_t^i \, d\bar{W}_t, \quad t \geq 0,
\end{equation}
where $( b^i, \sigma^i, \bar{\sigma}^i)$ are \( \mathbb{F} \)-progressively measurable processes taking values in \( \mathbb{R} \times \mathbb{R}_+ \times \mathbb{R}_+ \).

On one hand, for each $i=1,\dots,n$, the drift process $b^i$ and the idiosyncratic volatility process $\sigma^i$ are uncertain. Define the set of possibly realized drift process by
\begin{equation*}
\mathcal{B}^i=\left\{b^i:b^i\text{ is $\mathbb{F}$-progressively measurable, and $b^i\in U_b^i$, $\mathbb{P}\otimes \mathbb{L}$-a.e.}\right\},
\end{equation*}
where $U_b^i$ is a compact interval in $\mathbb{R}$, and $\mathbb{P}\otimes \mathbb{L}$ is the product measure of $\mathbb{P}$ and $\mathbb{L}$, where $\mathbb{L}$ is the Lebesgue measure; similarly, define the set of possibly realized idiosyncratic volatility process by
\begin{equation*}
\Sigma^i=\left\{\sigma^i:\sigma^i\text{ is $\mathbb{F}$-progressively measurable, and $\sigma^i\in U_\sigma^i$, $\mathbb{P}\otimes \mathbb{L}$-a.e.}\right\},
\end{equation*}
where $U_\sigma^i$ is a compact subset in $\mathbb{R}_+$, {but not necessarily convex.} 
For notational brevity, denote $b=\left(b^1,\dots,b^n\right)\in\mathcal{B}=\mathcal{B}^1\times\dots\times\mathcal{B}^n$,
with $U_b=U_b^1\times\dots\times U_b^n$, and $\sigma=\left(\sigma^1,\dots,\sigma^n\right)\in\Sigma=\Sigma^1\times\dots\times\Sigma^n$,
with $U_{\sigma}=U_{\sigma}^1\times\dots\times U_{\sigma}^n$. The larger the set $\mathcal{B}$ {(and respectively, $\Sigma$)} is, the more the drift {(and respectively, idiosyncratic volatility)} process is uncertain.

{O}n the other hand, for each $i=1,\dots,n$, the systematic volatility process $\bar{\sigma}^i$ is certain, $\mathbb{F}$-progressively measurable, and uniformly bounded.
Hence, in the remaining of this paper, the volatility uncertainty is interpreted as the uncertainty of idiosyncratic volatility process.

{The stock price model in \eqref{stock}, known as a \textit{systematic risk factor model}, captures both systematic and idiosyncratic risks, as outlined by Sharpe \cite{Sharpe_1963,Sharpe_1964}. The total risk of a risky stock $S^i$ consists of (i) \textit{systematic risk}, driven by the stock's co-movement with the market, modeled by the term $\bar{\sigma}_t^i \, d\bar{W}_t$ through the common Brownian component $\bar{W}_t$, and (ii) \textit{idiosyncratic risk}, independent of the market, modeled by the term $\sigma_t^i \, dW_t^i$ through independent Brownian components $W_t^i$, for $i = 1, \dots, n$.}

\begin{example}\label{example}
(Stochastic factor model)
The independent Brownian components $B_t=(B_t^1,\dots,B_t^{n})$, $t\geq 0$, cannot be directly traded through the stocks $(S^1,\dots, S^n)$.  A typical example is the following stochastic factor model: for each $i=1,\dots,n$, and for any $t\geq 0$,
\begin{equation*}
\frac{dS_t^i}{S_t^i}=b_t^idt+\sigma_t^idW_t^i+\bar{\sigma}^i(V_t)d\bar{W}_t,
\end{equation*}
with $V=(V^1,\dots,V^n)$ satisfying, for each $i=1,\dots,n$, and for any $t\geq 0$,
$$dV_t^i=\eta^i(V_t)dt+\kappa^i(V_t)dB_t^i+\bar{\kappa}^i(V_t)d\bar{W}_t,$$
where $\bar{\sigma}^i(\cdot), \eta^i(\cdot), \kappa^i(\cdot)$, and $\bar{\kappa}^i(\cdot)$ are some {bounded and measurable functions. 
When $n = 1$, this model is widely used in mathematical finance to capture
stochastic volatility. For the case of uncertain coefficients,
see \cite{Tevzadze_2013}; for earlier work with certain coefficients, see \cite{Zariphopoulou_2001, Zariphopoulou_2004}.}
\end{example}

A risk-averse and ambiguity-averse agent, who has an initial endowment $\xi\in\mathbb{R}_+$, can choose to consume,
and invest dynamically among the risk-free bond and the $n$ risky stocks in this physical financial market. Let $\pi_t=\left(\pi^1_t,\dots,\pi^n_t\right)$, $t\geq 0$, where, for each $i=1,\dots,n$, $\pi^i$ is a process representing the proportion of the agent's wealth in the $i$-th risky stock. Let $c_t$, $t\geq 0$, be a process representing the agent's consumption rate proportion of her wealth. Then, by self-financing, her wealth process $X_t^{\pi,c;b,\sigma}$, $t\geq 0$, solves, for any $t\geq 0$,
\begin{equation}
dX_t=X_t\left(\left(r+\sum_{i=1}^{n}\pi_t^i\left(b_t^i-r\right)-c_t\right)dt+\sum_{i=1}^{n}\pi_t^i\sigma_t^idW_t^i+\sum_{i=1}^{n}\pi_t^i\bar{\sigma}_t^id\bar{W}_t\right),
\label{wealth}
\end{equation}
with $X_0^{\pi,c;b,\sigma}=\xi$. Define the set of admissible investment and consumption strategies by
\begin{align*}
\mathcal{A}=\Bigg\{&\left(\pi_t,c_t\right),\;t\geq 0:\left(\pi,c\right)\text{ are }\mathbb{F}\text{-progressively measurable};\\&\left(\pi,c\right)\in\Pi\times\mathbb{R}_+,\;\mathbb{P}\otimes \mathbb{L}\text{-a.e.};\;\text{for any $t\geq 0$, }\int_{0}^{t}c_sds<\infty,\;\mathbb{P}\text{-a.s.}\Bigg\},
\end{align*}
where $\Pi$ is a convex and compact subset in $\mathbb{R}^n$ including the origin $0\in\mathbb{R}^n$.

In the sequel, for any times $t\geq 0$ and $T>t$, denote $\mathcal{A}_{\left[t,T\right)}$, $\mathcal{B}_{\left[t,T\right)}$, and $\Sigma_{\left[t,T\right)}$ respectively as the set of admissible investment and consumption strategies $\left(\pi,c\right)$, the set of possibly realized drift process $b$, and the set of possibly realized idiosyncratic volatility process $\sigma$, restricting in $\left[t,T\right)$; each element in $\mathcal{A}_{\left[t,T\right)}$, $\mathcal{B}_{\left[t,T\right)}$, and $\Sigma_{\left[t,T\right)}$, are respectively denoted by $\left(\pi,c\right)_{\left[t,T\right)}$, $b_{\left[t,T\right)}$, and $\sigma_{\left[t,T\right)}$. With a slight abuse of notation, for any time $t\geq 0$, $\left(\pi,c\right)_{\left[t,t\right)}$, $b_{\left[t,t\right)}$, and $\sigma_{\left[t,t\right)}$ are null, while $\mathcal{A}_{\left[t,t\right)}$, $\mathcal{B}_{\left[t,t\right)}$, and $\Sigma_{\left[t,t\right)}$ are null sets; if a mathematical object is said to be depending on $\left(\pi,c\right)_{\left[t,t\right)}$, $b_{\left[t,t\right)}$, or $\sigma_{\left[t,t\right)}$, the object is independent of it.

\begin{remark}
Note the dependence of the agent's wealth process $X^{\pi,c;b,\sigma}$ on, not only her choices of the investment and consumption strategies $\left(\pi,c\right)$, but also the market-realized drift process $b\in\mathcal{B}$ and idiosyncratic volatility process $\sigma\in\Sigma$. In particular, due to \eqref{wealth}, for any $t\geq 0$, the agent's wealth $X_t^{\pi,c;b,\sigma}$ depends on $\left(\pi,c\right)_{\left[0,t\right)}\in\mathcal{A}_{\left[0,t\right)}$ and $\left(b,\sigma\right)_{\left[0,t\right)}\in\left(\mathcal{B}\times\Sigma\right)_{\left[0,t\right)}$.
\end{remark}

\subsection{Robust Forward Investment and Consumption Preferences}\label{sub_section_2.2}
{
The risk-averse and ambiguity-averse agent is endowed with initial (time-$0$) investment and consumption preferences $U\left(x,0\right)$ and $U^c\left(C,0\right)$, which are non-decreasing and concave in $x\in\mathbb{R}_+$ and $C\in\mathbb{R}_+$, where $C=c\times x$ for $c\in\mathbb{R}_+$. In the physical financial market, her {\it robust forward investment and consumption preferences}, {\it with drift and volatility uncertainties}, and the associated optimal investment and consumption strategies, are defined as follows.}

\begin{definition} \label{def:forward_performance_drift_vol}
A pair of processes
\begin{equation*}
\{(U(\omega,x,t),U^c(\omega,C,t))\}_{\omega\in\Omega,x\in\mathbb{R}_+,C\in\mathbb{R}_+,t\geq 0}
\end{equation*}
is called robust forward investment and consumption preferences, with drift and volatility uncertainties, if they satisfy all of the following properties:
\begin{enumerate}
\item[(i)] for any $ x\in\mathbb{R}_+$, $ C\in\mathbb{R}_+$, and $t\geq 0$, $ \{U(\omega,x,t)\}_{\omega\in\Omega} $ and $ \{U^c(\omega,C,t)\}_{\omega\in\Omega} $ are $\mathcal{F}_t$-measurable;
\item[(ii)] for any $ \omega\in\Omega $ and $ t\geq 0 $, $ \{U(\omega,x,t)\}_{x\in\mathbb{R}_+} $ and $ \{U^c(\omega,C,t)\}_{C\in\mathbb{R}_+} $ are non-decreasing and concave;
\item[(iii)] for any $t\geq 0$, $\xi\in\mathcal{L}\left(\mathcal{F}_t;\mathbb{R}_+\right)$, and $T\geq t$,
\begin{equation}
\begin{aligned}
&\;U\left(\xi,t\right)=\esssup_{\left(\pi,c\right)\in\mathcal{A}}\essinf_{\left(b,\sigma\right)\in\mathcal{B}\times\Sigma}\\&\;\mathbb{E}\left[U\left(X_T^{\xi,t;\pi,c;b,\sigma},T\right)+\int_{t}^{T}U^c\left(c_sX_s^{\xi,t;\pi,c;b,\sigma},s\right)ds\vert\mathcal{F}_t\right],
\end{aligned}
\label{eq:forward_self_gen}
\end{equation}
where $\mathcal{L}\left(\mathcal{F}_t;\mathbb{R}_+\right)$ is the set of $\mathcal{F}_t$-measurable and $\mathbb{R}_+$-valued random variables, and $X^{\xi,t;\pi,c;b,\sigma}$ solves \eqref{wealth} with $X^{\xi,t;\pi,c;b,\sigma}_t=\xi$.
\end{enumerate}
If there exist a pair of forward investment and consumption strategies $\left(\pi^*,c^*\right)\in\mathcal{A}$ solving \eqref{eq:forward_self_gen}, it is called optimal and robust in the physical financial market.
\end{definition}

{We make some comments on the above definition. 
For any times $t\geq 0$ and $T\geq t$, 
 the essential supremum in \eqref{eq:forward_self_gen} is taken over all admissible investment and consumption strategies $(\pi, c) \in \mathcal{A}_{[t, T)}$, and the essential infimum is taken over all admissible drift and volatility coefficients $(b, \sigma) \in \mathcal{B}_{[t, T)} \times \Sigma_{[t, T)}$. The optimal and robust strategies $(\pi^*, c^*)$, if they exist, are required to be independent of the starting time $t$ and terminal time $T$, i.e., $(\pi^*, c^*) \in \mathcal{A}$. 

 Notably, the existence of worst-case coefficients for $(b, \sigma)$ that attain the infimum is not required; in fact, the infimum in \eqref{eq:forward_self_gen} may be achieved only in a randomized sense, through the use of a probabilistic mixture over admissible parameters.
} 

{
\subsection{Comparisons of Market Models and Preferences}

One may consider generalizing \eqref{stock} to a more general diffusion model including the systematic risk factor model as a special case, as in \cite{Chong_2024}. In such models, where drift and volatility are uncertain, the worst-case coefficients typically attain values at the boundaries of the uncertainty sets \( U_b \) and \( U_\sigma \). Consequently, the model reduces to the one with constant coefficients, rendering the market effectively complete.

A similar comparison arises in the context of optimal investment under static preferences. In \cite{YLZ_2017}, a general diffusion model with uncertain drift and volatility yields constant worst-case coefficients as the boundaries of the uncertainty sets, implying a complete market. In contrast, \cite{Matoussi_2015, Tevzadze_2013} consider uncertain volatility with known but \textit{stochastic} drift, leading to the worst-case volatility distribution and an incomplete market. Related works include \cite{Biagini_2017,Bordigoni_2007,Fouque_2016,Hernandez_2006,Lin_2014,Schied_2008}.
In our framework, we assume an incomplete market and uncertainty in both the drifts \( b^i \) and idiosyncratic volatilities \( \sigma^i \), while the systematic volatility \( \bar{\sigma}^i \) is \textit{stochastic} but known. As a result, the model represents a generic incomplete market.

Table \ref{tab:comparison} presents a comparison of our framework with key related work in the literature, highlighting differences in preferences, uncertainty, market completeness, and methods.} 

\begin{table}[ht]
\footnotesize
\centering
\setlength{\tabcolsep}{3pt}
\begin{tabular}{|p{2cm}|p{1.8cm}|p{1.8cm}|p{1.8cm}|p{1.8cm}|p{1.8cm}|}
\hline
 & Preference & Uncertainty & Market & Saddle Point & Method \\
\hline
Yang, Liang, and Zhou \cite{YLZ_2017} & Static & Drift \& Vol & Complete & Yes & PDE \\
\hline
Tevzadze, Toronjadze, and Uzunashvili \cite{Tevzadze_2013} & Static & Vol & Incomplete & No & Randomized PDE \\
\hline
Matoussi, Possama{\"i}, and Zhou \cite{Matoussi_2015} & Static & Vol & Incomplete & No & 2BSDE \\
\hline
Chong and Liang \cite{Chong_2024} & Forward & Drift \& Vol & Complete & Yes & PDE \\
\hline
K\"allblad, Ob{\l}{\'o}j, and Zariphopoulou \cite{Kallblad_2018} & Forward & Drift & Incomplete & Yes & Convex Duality \\
\hline
This paper & Forward & Drift \& Idiosyncratic Vol & Incomplete & No & Randomized BSDE \\
\hline
\end{tabular}
\caption{Comparison of market models and preferences.}
\label{tab:comparison}
\end{table}

\section{Robust Randomized Forward Investment and Consumption Preferences}\label{sec:3}

{As discussed in the introduction, the incompleteness of the market implies that the worst-case volatility may not exist as a saddle point. To address this, we introduce a randomization of volatility by considering the induced probability measure over the uncertainty set, thereby constructing an auxiliary market. Within this auxiliary market, the agent formulates robust randomized forward investment and consumption preferences.}       

We first rewrite the physical financial market in terms of measure-valued processes.
To this end, for each $i=1,\dots,n$, define the set of Dirac measure-valued processes on $U_{\sigma}^{i}$, by
\begin{align*}
\left(\mathcal{P}^i\right)'\left(U_{\sigma}^{i}\right)=&\left\{m^i_t,\;t\geq 0:m^i\text{ is $\mathbb{F}$-progressively measurable, and }\right.\\&\;\left.\quad\quad\quad\quad \quad \;m^i=\delta^i_{u^i},\text{ for some }u^i\in U_{\sigma}^{i},\;\mathbb{P}\otimes \mathbb{L}\text{-a.e.}\right\},
\end{align*}
where, for any $u^i\in U_{\sigma}^i$, and $B\in\mathbb{B}\left(U_{\sigma}^i\right)$, which is the Borel $\sigma$-algebra on $U_{\sigma}^i$, $\delta^i_{u^i}\left(B\right)=1$ if $u^i\in B$, while $\delta^i_{u^i}\left(B\right)=0$ if $u^i\notin B$. Note that a possible realization $u^i\in U_{\sigma}^i$, in $\left(\mathcal{P}^i\right)'\left(U_{\sigma}^{i}\right)$, is random and depends on time. For notational brevity, denote $\delta=\left(\delta^1,\dots,\delta^n\right)\in\mathcal{P}'\left(U_{\sigma}\right)=\left(\mathcal{P}^1\right)'\left(U_{\sigma}^1\right)\times\dots\times\left(\mathcal{P}^n\right)'\left(U_{\sigma}^n\right)$. For any $\left(\pi,c\right)\in\mathcal{A}$ and $\left(b,\delta\right)\in\mathcal{B}\times\mathcal{P}'\left(U_{\sigma}\right)$, define a process $X_t^{\pi,c;b,\delta}$, $t\geq 0$, which satisfies, for any $t\geq 0$,
\begin{equation}
\begin{aligned}
dX_t=&\;X_t\left(\left(r+\sum_{i=1}^{n}\pi_t^i\left(b_t^i-r\right)-c_t\right)dt\right.\\&\left.+\sum_{i=1}^{n}\pi_t^i\sqrt{\int_{U_{\sigma}^i}
\tilde{u}^2\delta^i_{u^i}\left(d\tilde{u}\right)}dW_t^i+\sum_{i=1}^{n}\pi_t^i\bar{\sigma}_t^id\bar{W}_t\right),
\end{aligned}
\label{wealth_random}
\end{equation}
with $X_0^{\pi,c;b,\delta}=\xi$. 

Fix any $\left(\pi,c\right)\in\mathcal{A}$ and $b\in\mathcal{B}$. For any $\sigma\in\Sigma$, there exists a Dirac measure-valued process $\delta\in\mathcal{P}'\left(U_{\sigma}\right)$ such that $X^{\pi,c;b,\sigma}$ and $X^{\pi,c;b,\delta}$ are indistinguishable, which is given by $\delta^i_{u^i}=\delta^i_{\sigma^i}$, for each $i=1,\dots,n$. Similarly, for any $\delta\in\mathcal{P}'\left(U_{\sigma}\right)$, there exists an idiosyncratic volatility process $\sigma\in\Sigma$ such that $X^{\pi,c;b,\delta}$ and $X^{\pi,c;b,\sigma}$ are indistinguishable, which is given by $\sigma^i=u^i$, for each $i=1,\dots,n$. Therefore, the wealth process of the agent $X^{\pi,c;b,\sigma}$, for some market-realized idiosyncratic volatility process $\sigma\in\Sigma$, can be identified by the process $X^{\pi,c;b,\delta}$, which depends on the Dirac measure-valued process $\delta\in\mathcal{P}'\left(U_{\sigma}\right)$, and vice versa. Thus, the condition (iii) in Definition \ref{def:forward_performance_drift_vol} is equivalent to that, for any $t\geq 0$, $\xi\in\mathcal{L}\left(\mathcal{F}_t;\mathbb{R}_+\right)$, and $T\geq t$,
\begin{equation*}
\begin{aligned}
&\;U\left(\xi,t\right)=\esssup_{\left(\pi,c\right)\in\mathcal{A}}\essinf_{\left(b,\delta\right)\in\mathcal{B}\times\mathcal{P}'\left(U_{\sigma}\right)}\\&\;\mathbb{E}\left[U\left(X_T^{\xi,t;\pi,c;b,\delta},T\right)+\int_{t}^{T}U^c\left(c_sX_s^{\xi,t;\pi,c;b,\delta},s\right)ds\vert\mathcal{F}_t\right],
\end{aligned}
\end{equation*}
where $X^{\xi,t;\pi,c;b,\delta}$ solves \eqref{wealth_random} with $X^{\xi,t;\pi,c;b,\delta}_t=\xi$. For any $t\geq 0$ and $T\geq t$, denote $\mathcal{P}'\left(U_{\sigma}\right)_{\left[t,T\right)}$ as the set of Dirac measure-valued process $\delta$ restricting in $\left[t,T\right)$; each element in $\mathcal{P}'\left(U_{\sigma}\right)_{\left[t,T\right)}$ is denoted by $\delta_{\left[t,T\right)}$.

Due to such equivalence, the financial market is {\it still} the physical one without any randomization on the idiosyncratic volatility process.

\subsection{The Auxiliary Market}
For each $i=1,\dots,n$, denote by $\mathcal{P}^{i}\left(U_{\sigma}^{i}\right)$
the set of $\mathbb{F}$-progressively measurable Borel probability measure-valued processes $m^i_t$, $t\geq 0$, on the set $U_{\sigma}^{i}$; also, for any $\omega\in\Omega$ and $t\geq 0$, denote by $\mathcal{P}^{i}\left(U_{\sigma}^{i}\right)_{\omega,t}$ the restriction of $\mathcal{P}^{i}\left(U_{\sigma}^{i}\right)$ at $\left(\omega,t\right)$, which is in fact the set of Borel probability measures on the set $U_{\sigma}^{i}$. Obviously, both $\mathcal{P}^{i}\left(U_{\sigma}^{i}\right)$ and $\mathcal{P}^{i}\left(U_{\sigma}^{i}\right)_{\omega,t}$, for $\omega\in\Omega$ and $t\geq 0$, are convex. Since $U_{\sigma}^{i}$ is compact,
for any $\omega\in\Omega$ and $t\geq 0$, $\mathcal{P}^{i}\left(U_{\sigma}^{i}\right)_{\omega,t}$ is compact under the weak convergence and topology on $U_{\sigma}^{i}$. For notational brevity, denote $m=\left(m^1,\dots,m^n\right)\in\mathcal{P}\left(U_{\sigma}\right)=\mathcal{P}^1\left(U_{\sigma}^{1}\right)\times\dots\times\mathcal{P}^n\left(U_{\sigma}^{n}\right)$. Clearly, $\mathcal{P}'\left(U_{\sigma}\right)\subseteq\mathcal{P}\left(U_{\sigma}\right)$.

In the auxiliary financial market, the stock price process depends on the Borel probability measure-valued process $m\in\mathcal{P}\left(U_{\sigma}\right)$ in place of the idiosyncratic volatility process $\sigma\in\Sigma$: for each $i=1,\dots,n$, and for any $t\geq 0$,
\begin{equation*}
\frac{dS_t^i}{S_t^i}=b_t^idt+\sqrt{\int_{U_{\sigma}^i}u^2m^i_t\left(du\right)}dW_t^i+\bar{\sigma}_t^id\bar{W}_t.
\end{equation*}
Therefore, for an agent living in this auxiliary market, her wealth process $X_t^{\pi,c;b,m}$, $t\geq 0$, satisfies, for any $t\geq 0$,
\begin{equation}
\begin{aligned}
dX_t=&\;X_t\left(\left(r+\sum_{i=1}^{n}\pi_t^i\left(b_t^i-r\right)-c_t\right)dt\right.\\&\left.+\sum_{i=1}^{n}\pi_t^i\sqrt{\int_{U_{\sigma}^i}u^2m^i_t\left(du\right)}dW_t^i+\sum_{i=1}^{n}\pi_t^i\bar{\sigma}_t^id\bar{W}_t\right),
\end{aligned}
\label{wealth_random_2}
\end{equation}
with $X_0^{\pi,c;b,m}=\xi$, for any $\left(\pi,c\right)\in\mathcal{A}$ and $\left(b,m\right)\in\mathcal{B}\times\mathcal{P}\left(U_{\sigma}\right)$. Denote the self-evident notations $m_{\left[t,T\right)}\in\mathcal{P}\left(U_{\sigma}\right)_{\left[t,T\right)}$, for any times $t\geq 0$ and $T\geq t$, and obviously, $\mathcal{P}\left(U_{\sigma}\right)_{\omega,t}$ the set of Borel probability measures on the set $U_{\sigma}$. When $m\in\mathcal{P}'\left(U_{\sigma}\right)$, \eqref{wealth}, \eqref{wealth_random}, and \eqref{wealth_random_2} coincide.

\begin{remark}
Note that the Borel probability measure-valued process  $m\in\mathcal{P}\left(U_{\sigma}\right)$ is $\mathbb{F}$-progressively measurable. That is, the randomization via $m$ is not based on any exogenous randomness imposing to the physical financial market. The randomization is thus {\it endogenous} in this paper, without enlarging the filtration $\mathbb{F}$, which is indeed in line with \cite{Tevzadze_2013} for static preferences.
\end{remark}

\subsection{Robust Randomized Forward Preferences}
{To account for uncertainty modeled by probability measure-valued processes in this auxiliary financial market, we generalize Definition \ref{def:forward_performance_drift_vol} by incorporating the probability measure-valued process. The robust randomized forward investment and consumption preferences are defined as follows.}

\begin{definition} \label{def:forward_performance_drift_vol_random}
A pair of processes
\begin{equation*}
\{(U(\omega,x,t;m_{\left[0,t\right)}),U^c(\omega,C,t))\}_{\omega\in\Omega,x\in\mathbb{R}_+,C\in\mathbb{R}_+,t\geq 0,m_{\left[0,t\right)}\in\mathcal{P}\left(U_{\sigma}\right)_{\left[0,t\right)}}
\end{equation*}
is called robust randomized forward investment and consumption preferences, with drift and volatility uncertainties, if they satisfy all of the following properties:
\begin{enumerate}
\item[(i)] for any $ x\in\mathbb{R}_+$, $ C\in\mathbb{R}_+$, $t\geq 0$, and $m_{\left[0,t\right)}\in\mathcal{P}\left(U_{\sigma}\right)_{\left[0,t\right)}$, $ \{U(\omega,x,t;m_{\left[0,t\right)})\}_{\omega\in\Omega} $ and $ \{U^c(\omega,C,t)\}_{\omega\in\Omega} $ are $\mathcal{F}_t$-measurable;
\item[(ii)] for any $ \omega\in\Omega $, $t\geq 0$, and $m_{\left[0,t\right)}\in\mathcal{P}\left(U_{\sigma}\right)_{\left[0,t\right)}$, $ \{U(\omega,x,t;m_{\left[0,t\right)})\}_{x\in\mathbb{R}_+} $ and $ \{U^c(\omega,C,t)\}_{C\in\mathbb{R}_+} $ are non-decreasing and concave;
\item[(iii)] for any $t\geq 0$, $\xi\in\mathcal{L}\left(\mathcal{F}_t;\mathbb{R}_+\right)$, $m_{\left[0,t\right)}\in\mathcal{P}\left(U_{\sigma}\right)_{\left[0,t\right)}$, and $T\geq t$,
\begin{equation}
\begin{aligned}
&\;U\left(\xi,t;m_{\left[0,t\right)}\right)=\esssup_{\left(\pi,c\right)\in\mathcal{A}}\;\essinf_{\left(b,m\right)_{\left[t,T\right)}\in\left(\mathcal{B}\times\mathcal{P}\left(U_{\sigma}\right)\right)_{\left[t,T\right)}}\\&\;\mathbb{E}\left[U\left(X_T^{\xi,t;\pi,c;b,m},T;m_{\left[0,T\right)}\right)+\int_{t}^{T}U^c\left(c_sX_s^{\xi,t;\pi,c;b,m},s\right)ds\vert\mathcal{F}_t\right],
\end{aligned}
\label{eq:forward_self_gen_random}
\end{equation}
where $X^{\xi,t;\pi,c;b,m}$ solves \eqref{wealth_random_2} with $X^{\xi,t;\pi,c;b,m}_t=\xi$, $m_{\left[0,T\right)}=m_{\left[0,t\right)}\oplus m_{\left[t,T\right)}$, and $\oplus$ is the time-pasting binary operator.
\end{enumerate}
If there exis{t a pair of} forward investment and consumption strategies $\left(\pi^*,c^*\right)\in\mathcal{A}$ solving \eqref{eq:forward_self_gen_random}, it is called optimal and robust in the auxiliary financial market.
\end{definition}

{We make some comments on the above definition. 
The introduction of Borel probability measure-valued processes \( m_{[0,t)} \in \mathcal{P}(U_\sigma)_{[0,t)} \) extends the space of coefficients to their induced probability measures.
This randomization ensures that the essential infimum in \eqref{eq:forward_self_gen_random} achieves a saddle point as a probability distribution in the auxiliary market.
The robust randomized forward investment preference \( U(\omega, x, t; m_{[0,t)}) \) at time \( t \) depends on the history of the Borel measure-valued process \( m_{[0,t)} \). This dependence introduces randomization, as the preference adapts to the evolving distribution of market parameters.}

\section{Robust Randomized CRRA Forward Preferences}\label{sec:4}

{ In this section, we construct a class of robust randomized CRRA forward investment and consumption preferences. Specifically, we construct $\mathbb{F}$-progressively measurable processes $K_t$ and $\lambda_t$, for $t \geq 0$, such that for any $x \in \mathbb{R}_+$, $C \in \mathbb{R}_+$, $t \geq 0$, and $m_{[0,t)} \in \mathcal{P}(U_{\sigma})_{[0,t)}$, the preferences are given by
\[
U(x,t; m_{[0,t)}) = \frac{x^{\kappa}}{\kappa} e^{K_t}, \quad \text{and} \quad U^c(C,t) = \frac{C^{\kappa}}{\kappa} \lambda_t,
\]
where $\kappa \in (0,1)$ denotes the agent's risk aversion parameter. In particular, at the initial time $t=0$, the agent’s investment and consumption preferences are respectively given by 
\[
U(x,0) = \frac{x^{\kappa}}{\kappa} e^{K_0}, \quad \text{and} \quad U^c(C,0) = \frac{C^{\kappa}}{\kappa} \lambda_0.
\]

The structure of these robust randomized forward preferences is inspired by the classical framework involving mathematically tractable utility functions (see, for example, \cite{HU0} and \cite{Cheridito_2011}). Since $K$ is $\mathbb{F}$-progressively measurable and independent of the state variable $x$, the resulting forward investment preference exhibits non-zero volatility, capturing the essential feature of market incompleteness. This stands in contrast to the zero-volatility case studied in \cite{Chong_2024} with $K$ being deterministic, which effectively corresponds to a complete market under the worst-case model parameters.

The process $\lambda$, also $\mathbb{F}$-progressively measurable and independent of the state variable $C$, represents the agent’s evolving beliefs about future consumption preferences. While we will later derive a BSDE representation for the process $K$, our results show that no such dynamics are required for $\lambda$—other than being uniformly bounded and satisfying a structural condition. Specifically, to ensure the joint existence of forward investment and consumption preferences, it is sufficient that the forward consumption preference is dominated by the initial investment preference. A broad class of processes $\lambda$ satisfies this requirement. For example, if $\lambda$ is dominated by a non-increasing function of time, it can naturally be interpreted as a stochastic discount factor that reflects declining marginal utility from consumption over time. See Example~\ref{example_lambda} for further illustration.}

\subsection{Saddle Point of Randomized Hamiltonian {by Convexification}}\label{sec:saddle}
{T}he saddle point for the Hamiltonian does not generally exist in the physical market. This section is dedicated to demonstrating that the saddle point for the randomized Hamiltonian \textit{does exist} in the auxiliary market. Consequently, in Section \ref{sec:repre}, the saddle point and its corresponding value are used to construct the robust randomized forward preferences, along with the optimal and robust strategies.
The motivation for studying this form of the randomized Hamiltonian stems from the approach that the Brownian {components} $B^i$, for $i=1,\dots,n$, which cannot be fully hedged by the stocks in the market, are also endogenously randomized in the next section. This randomization facilitates the construction of preferences and the solution of the associated strategies.

For any $\omega\in\Omega$, $t\geq 0$, $z\in\mathbb{R}^n$, $\bar{z}\in\mathbb{R}$, define a {Hamiltonian} function, $H\left(\omega,t,z,\bar{z};\cdot;\cdot,\cdot\right):\Pi\times U_b\times\mathcal{P}\left(U_{\sigma}\right)_{\omega,t}\rightarrow\mathbb{R}$, as follows: for any $\left(x_{\pi};x_{b},x_{m}\right)\in\Pi\times U_b\times\mathcal{P}\left(U_{\sigma}\right)_{\omega,t}$,
\begin{equation}
\begin{aligned}
H\left(\omega,t,z,\bar{z};x_{\pi};x_{b},x_{m}\right)=&\;-\frac{1}{2}\kappa\left(1-\kappa\right)\sum_{i=1}^{n}\left(x_{\pi}^{i}\right)^2\left(\int_{U_{\sigma}^{i}}u^2x_m^i\left(du\right)+\left(\bar{\sigma}_t^i\right)^2\right)\\&\;+\kappa\sum_{i=1}^{n}x_{\pi}^i\left(x_b^i-r+\rho^iz^i\int_{U_{\sigma}^i}ux_m^i\left(du\right)+\bar{\sigma}_t^i\bar{z}\right)\\&\;+\frac{1}{2}\left(\sum_{i=1}^{n}\left(z^i\right)^2+\left(\bar{z}\right)^2\right)+\kappa r.
\end{aligned}
\label{eq:H}
\end{equation}
In particular, when $x_{m}\in\mathcal{P}'\left(U_{\sigma}\right)_{\omega,t}$, which is the set of Dirac measures on the set $U_{\sigma}$, the {Hamiltonian} function can be identified as follows: for any $\omega\in\Omega$, $t\geq 0$, $z\in\mathbb{R}^n$, $\bar{z}\in\mathbb{R}$, and for any $\left(x_{\pi};x_{b},x_{\sigma}\right)\in\Pi\times U_b\times U_{\sigma}$,
\begin{equation}
\begin{aligned}
H\left(\omega,t,z,\bar{z};x_{\pi};x_{b},x_{\sigma}\right)=&\;-\frac{1}{2}\kappa\left(1-\kappa\right)\sum_{i=1}^{n}\left(x_{\pi}^{i}\right)^2\left(\left(x_{\sigma}^{i}\right)^2+\left(\bar{\sigma}_t^i\right)^2\right)\\&\;+\kappa\sum_{i=1}^{n}x_{\pi}^i\left(x_b^i-r+\rho^iz^ix_{\sigma}^{i}+\bar{\sigma}_t^i\bar{z}\right)\\&\;+\frac{1}{2}\left(\sum_{i=1}^{n}\left(z^i\right)^2+\left(\bar{z}\right)^2\right)+\kappa r.
\end{aligned}
\label{eq:H_canonical}
\end{equation}
{The two Hamiltonians are related through convexification, where the randomized Hamiltonian is the expectation of the deterministic Hamiltonian over the probability measure $x_m$:
\begin{equation*}
H(\omega, t, z, \bar{z}; x_{\pi}, x_b, x_m) = \int_{U_{\sigma}} H(\omega, t, z, \bar{z}; x_{\pi}, x_b, u) \, x_m(du).
\end{equation*}

Note that the correlation $\rho^i$ in the Hamiltonian \eqref{eq:H_canonical} introduces the cross-term $\kappa \rho^i z^i x_{\pi}^i x_{\sigma}^i$, which, together with the negative quadratic term $-\frac{1}{2} \kappa (1 - \kappa) (x_{\pi}^i)^2 (x_{\sigma}^i)^2$, renders the Hamiltonian non-convex in $x_{\sigma}^i$ and nonlinear in $(x_{\sigma}^i)^2$. In contrast, the Hamiltonian \eqref{eq:H} is linear in the probability measure $x_m$.}

{
\begin{lemma}\label{lemma:H_saddle_point} 
Fix any $\omega \in \Omega$, $t \geq 0$, $z \in \mathbb{R}^n$, and $\bar{z} \in \mathbb{R}$. Let $H(\omega, t, z, \bar{z}; \cdot; \cdot, \cdot)$ be the Hamiltonian defined in \eqref{eq:H}. Define the projection function $\text{Proj}_{\Pi^i}(a) = \argmin_{b \in \Pi^i} |a - b|$ and the distance function $\text{dist}(\Pi^i, a) = \min_{b \in \Pi^i} |a - b|$, for any $a \in \mathbb{R}$. The Hamiltonian $H$ satisfies the following properties:
\begin{enumerate}
\item[(i)] Maximin Equality:
\begin{align*}
&\;\sup_{x_{\pi}\in\Pi}\inf_{\left(x_{b},x_{m}\right)\in U_b\times\mathcal{P}\left(U_{\sigma}\right)_{\omega,t}}H\left(\omega,t,z,\bar{z};x_{\pi};x_{b},x_{m}\right)\\=&\;\inf_{\left(x_{b},x_{m}\right)\in U_b\times\mathcal{P}\left(U_{\sigma}\right)_{\omega,t}}\sup_{x_{\pi}\in\Pi}H\left(\omega,t,z,\bar{z};x_{\pi};x_{b},x_{m}\right).
\end{align*}
\item[(ii)] Saddle-Point Existence: There exists a saddle point $(x_{\pi}^*, x_b^*, x_m^*) \in \Pi \times U_b \times \mathcal{P}(U_{\sigma})_{\omega, t}$, dependent on $(\omega, t, z, \bar{z})$, such that for all $x_{\pi} \in \Pi$ and $(x_b, x_m) \in U_b \times \mathcal{P}(U_{\sigma})_{\omega, t}$,
\begin{align*}
&\;H\left(\omega,t,z,\bar{z};x_{\pi};x_{b}^*,x_{m}^*\right)\\&\;\quad\quad\quad\quad\leq H\left(\omega,t,z,\bar{z};x_{\pi}^*;x_{b}^*,x_{m}^*\right)\leq H\left(\omega,t,z,\bar{z};x_{\pi}^*;x_{b},x_{m}\right).
\end{align*}
\item[(iii)] { Characterization}: The saddle point is given by $x_{\pi}^{i,*} = x_{\pi}^{i,*}(x_b^{i,*}, x_m^{i,*})$, for $i = 1, \ldots, n$, where
\begin{equation}
x_{\pi}^{i,*}\left(x_{b}^{i},x_{m}^{i}\right)=\text{Proj}_{\Pi^i}\left(\frac{x_b^{i}-r+\rho^iz^i\int_{U_{\sigma}^i}ux_m^{i}\left(du\right)+\bar{\sigma}_t^i\bar{z}}{\left(1-\kappa\right)\left(\int_{U_{\sigma}^i}u^2x_m^{i}\left(du\right)+\left(\bar{\sigma}_t^i\right)^2\right)}\right),
\label{eq:saddle_point_pi}
\end{equation}
\begin{equation}
\begin{aligned}
\left(x_{b}^{*},x_{m}^{*}\right)=&\;\argmin_{\left(x_{b},x_{m}\right)\in U_b\times\mathcal{P}\left(U_{\sigma}\right)_{\omega,t}}\sup_{x_{\pi}\in\Pi}H\left(\omega,t,z,\bar{z};x_{\pi};x_{b},x_{m}\right)\\=&\;\argmin_{\left(x_{b},x_{m}\right)\in U_b\times\mathcal{P}\left(U_{\sigma}\right)_{\omega,t}}H\left(\omega,t,z,\bar{z};x_{\pi}^{*}\left(x_{b},x_{m}\right);x_{b},x_{m}\right),
\end{aligned}
\label{eq:saddle_point_b_m}
\end{equation}
in which
\begin{equation}
\begin{aligned}
&\;H\left(\omega,t,z,\bar{z};x_{\pi}^{*}\left(x_{b},x_{m}\right);x_{b},x_{m}\right)\\=&\;\sum_{i=1}^{n}\left(-\frac{1}{2}\kappa\left(1-\kappa\right)\left(\int_{U_{\sigma}^{i}}u^2x_m^i\left(du\right)+\left(\bar{\sigma}_t^i\right)^2\right)\right.\\&\;\quad\quad\quad\quad\left.\times\text{dist}^2\left\{\Pi^i,\frac{x_b^{i}-r+\rho^iz^i\int_{U_{\sigma}^i}ux_m^{i}\left(du\right)+\bar{\sigma}_t^i\bar{z}}{\left(1-\kappa\right)\left(\int_{U_{\sigma}^i}u^2x_m^{i}\left(du\right)+\left(\bar{\sigma}_t^i\right)^2\right)}\right\}\right.\\&\;\quad\quad\;\left.+\frac{1}{2}\frac{\kappa}{1-\kappa}\frac{\left(x_b^{i}-r+\rho^iz^i\int_{U_{\sigma}^i}ux_m^{i}\left(du\right)+\bar{\sigma}_t^i\bar{z}\right)^2}{\int_{U_{\sigma}^i}u^2x_m^{i}\left(du\right)+\left(\bar{\sigma}_t^i\right)^2}\right)\\&\;+\frac{1}{2}\sum_{i=1}^{n}\left(z^i\right)^2+\frac{1}{2}\left(\bar{z}\right)^2+\kappa r.
\end{aligned}
\label{eq:H_sub_optimal}
\end{equation}
The saddle value function is
\begin{equation}
\begin{aligned}
H^*\left(\omega,t,z,\bar{z}\right)=&\;\sup_{x_{\pi}\in\Pi}\inf_{\left(x_{b},x_{m}\right)\in U_b\times\mathcal{P}\left(U_{\sigma}\right)_{\omega,t}}H\left(\omega,t,z,\bar{z};x_{\pi};x_{b},x_{m}\right)\\=&\;\inf_{\left(x_{b},x_{m}\right)\in U_b\times\mathcal{P}\left(U_{\sigma}\right)_{\omega,t}}\sup_{x_{\pi}\in\Pi}H\left(\omega,t,z,\bar{z};x_{\pi};x_{b},x_{m}\right)\\=&\;H\left(\omega,t,z,\bar{z};x_{\pi}^*;x_{b}^*,x_{m}^*\right)\\=&\;\inf_{\left(x_{b},x_{m}\right)\in U_b\times\mathcal{P}\left(U_{\sigma}\right)_{\omega,t}}H\left(\omega,t,z,\bar{z};x_{\pi}^{*}\left(x_{b},x_{m}\right);x_{b},x_{m}\right)\\=&\;H\left(\omega,t,z,\bar{z};x_{\pi}^{*}\left(x_{b}^*,x_{m}^*\right);x_{b}^*,x_{m}^*\right).
\end{aligned}
\label{eq:saddle_valeu_equivalence}
\end{equation}
\end{enumerate}
\end{lemma}}


\subsection{Representation by {Randomized} Infinite-Horizon BSDE and ODE}\label{sec:repre}
We will utilize {a family of randomized} infinite-horizon BSDEs to represent the robust randomized CRRA forward investment and consumption preferences, accounting for drift and volatility uncertainties, along with the corresponding optimal and robust forward investment and consumption strategies.

To begin with, we express \( B^i \) as follows: for any \( t \geq 0 \),
\[
B^i_t = \rho^i W^i_t + \sqrt{1 - (\rho^i)^2} W^{i+n}, \quad \text{for } i = 1, \dots, n,
\]
where \( W^{i+n} \) is one of the independent Brownian components that are independent of \( (W^1, \dots, W^n, \bar{W}) \). Given that the correlation coefficients \( \rho^i \), for \( i = 1, \dots, n \), are constant, the filtration \( \mathbb{F} \) can be regarded as generated by \( (W^1, \dots, {W^n, \bar{W}, W^{1+n}}, \dots, W^{2n}) \) after augmentation.

Inspired by \cite{Tevzadze_2013} for static preferences, the Brownian {components} \( B^i \) for \( i = 1, \dots, n \) are endogenously randomized as follows: for any \( m \in \mathcal{P}(U_{\sigma}) \), for each \( i = 1, \dots, n \), and for any \( t \geq 0 \),

\begin{equation}\label{new_BM}
B_{t}^{i,m^i} = \int_{0}^{t} \rho_s^{i,m^i} dW_s^i + \int_{0}^{t} \sqrt{1 - (\rho_s^{i,m^i})^2} dW_s^{i+n}.
\end{equation}
Here, the random correlation coefficient is defined as
\[
\rho_t^{i,m^i} = \frac{\int_{U_{\sigma}^i} u m_t^i(du)}{\sqrt{\int_{U_{\sigma}^i} u^2 m_t^i(du)}} \rho^i,
\]
which is uniformly bounded in \( [-1, 1] \) by Jensen's inequality and is \( \mathbb{F} \)-progressively measurable. According to Lévy's characterization theorem, for any \( m \in \mathcal{P}(U_{\sigma}) \), the process \( B_t^m = (B_{t}^{1,m^1}, \dots, B_{t}^{n,m^n}) \), \( t \geq 0 \), constitutes an \( n \)-dimensional independent Brownian motion on the filtered probability space \( (\Omega, \mathcal{F}, \mathbb{F}, \mathbb{P}) \), since both stochastic integrals in \eqref{new_BM} are \( \mathbb{F} \)-martingales, and $\langle  B^{i,m^i}_{\cdot}, B^{j,m^j}_{\cdot} \rangle_t = \delta^{ij}t$.

For each \( i = 1, \dots, n \), \( W^i \) and \( B^{i,m^i} \) are correlated such that 
\[
\langle W^i_{\cdot}, B^{i,m^i}_{\cdot} \rangle_t = \int_{0}^{t} \rho_s^{i,m^i} ds, \quad \text{for } t \geq 0,
\]
while \(W^j,\ \bar{W} \) and \( B^{i,m^i} \) for $i\neq j$ remain independent. When \( m \in \mathcal{P}'(U_{\sigma}) \), for each \( i = 1, \dots, n \), and for any \( t \geq 0 \), \( \rho^{i,m^i}_t = \rho^i \), thus yielding \( B_t^{m} = B_t \). It is also important to note that for any \( t \geq 0 \), \( B_t^m \) depends on \( m \in \mathcal{P}(U_{\sigma}) \) solely through \( m_{[0,t)} \in \mathcal{P}(U_{\sigma})_{[0,t)} \).

\begin{proposition}\label{prop:power_infinite_BSDE_existence_drift_vol}
For any $\rho>0$ and $m\in\mathcal{P}\left(U_{\sigma}\right)$, consider the {randomized} infinite-horizon BSDE that, for any $t\geq 0$,
\begin{equation}
dY_t=-\left(H^*\left(t,Z_t,\bar{Z}_t\right)-\rho Y_t\right)dt+\sum_{i=1}^{n}Z_t^{i}dB^{i,m^i}_t+\bar{Z}_td\bar{W}_t,
\label{eq:power_infinite_time_horizon_BSDE_drift_vol}
\end{equation}
where $H^*\left(\cdot,\cdot,\cdot,\cdot\right)$ is the saddle value function in \eqref{eq:saddle_valeu_equivalence} of Lemma \ref{lemma:H_saddle_point}, and $Z_t=\left(Z_t^{1},\dots,Z_t^{n}\right)$, $t\geq 0$. Then, for any $\rho>0$ and $m\in\mathcal{P}\left(U_{\sigma}\right)$, the infinite-horizon BSDE \eqref{eq:power_infinite_time_horizon_BSDE_drift_vol} admits the unique solution $\left(Y^m,Z^m,\bar{Z}^m\right) $, such that $ Y^m $ is $ \mathbb{F} $-progressively measurable, uniformly bounded, and continuous, $\mathbb{P}$-a.s., while $ \left(Z^m,\bar{Z}^m\right)\in\mathcal{L}^{2,-2\rho}[0,\infty) $, i.e. $Z^{1,m},\dots,Z^{n,m},\bar{Z}^m$ are $\mathbb{F}$-progressively measurable, and $\mathbb{E}\left[\int_{0}^{\infty}e^{-2\rho t}\left(\sum_{i=1}^{n}\left(Z^{i,m}_t\right)^2+\left(\bar{Z}^m_t\right)^2\right)dt\right]<\infty$; in particular, for any $ t\geq 0 $, $\mathbb{E}\left[\int_{0}^{t}\left(\sum_{i=1}^{n}\left(Z^{i,m}_s\right)^2+\left(\bar{Z}^m_s\right)^2\right)ds\right]<\infty$, which implies that $ \int_{0}^{t}\left(\sum_{i=1}^{n}\left(Z^{i,m}_s\right)^2+\left(\bar{Z}^m_s\right)^2\right)ds<\infty $, $ \mathbb{P} $-a.s.
\end{proposition}

\begin{remark}
Note that the unique solution $\left(Y^m,Z^m,\bar{Z}^m\right)$ of the infinite-horizon BSDE \eqref{eq:power_infinite_time_horizon_BSDE_drift_vol} depends on the Borel probability measure-valued process $m\in\mathcal{P}\left(U_{\sigma}\right)$, only via the Brownian motion $B^m$ in the equation. For any fixed $t\geq 0$, $\left(Y^m_t,Z^m_t,\bar{Z}^m_t\right)$ depends on $m\in\mathcal{P}\left(U_{\sigma}\right)$ only through $m_{\left[0,t\right)}\in\mathcal{P}\left(U_{\sigma}\right)_{\left[0,t\right)}$, but not via $x_{m}^*\in\mathcal{P}\left(U_{\sigma}\right)_{\omega,t}$ in the saddle point for the randomized Hamiltonian $H$ at the time $t$; recall that $\mathcal{P}\left(U_{\sigma}\right)_{\left[0,t\right)}$ is the set of Borel probability measure-valued processes restricting in $\left[0,t\right)$, while $\mathcal{P}\left(U_{\sigma}\right)_{\omega,t}$ is the set of Borel probability measures.
\end{remark}

\begin{theorem}\label{thm:power_forward_utility_consumption_drift_vol}
For any $\rho>0$ and $m\in\mathcal{P}\left(U_{\sigma}\right)$, let $\left(Y^m,Z^m,\bar{Z}^m\right)$ be the solution of the {randomized} infinite-horizon BSDE \eqref{eq:power_infinite_time_horizon_BSDE_drift_vol}; let $g^m_t$, $t\geq 0$, be the solution of the following ODE: for any $t\geq 0$,
\begin{equation}
dg_t=\left(\left(1-\kappa\right)\lambda_t^{\frac{1}{1-\kappa}}e^{-\frac{Y^m_t}{1-\kappa}}e^{\frac{g_t}{1-\kappa}}+\rho Y^m_t\right)dt,
\label{eq:g_t_nonlinear_ODE_drift_vol}
\end{equation}
where the {uniformly bounded and $\mathbb{F}$-progressively measurable process} 
$\lambda_t$, $t\geq 0$, satisfies the condition that, for any $t\geq 0$,
\begin{equation}
e^{-\frac{g^m_0}{1-\kappa}}>\int_{0}^{t}e^{\frac{1}{1-\kappa}\left(\rho\int_{0}^{s}Y^m_udu-Y^m_s\right)}\lambda_s^{\frac{1}{1-\kappa}}ds.
\label{eq:ODE_condition_drift_vol}
\end{equation}
The following two assertions hold.

(i) The pair of processes
\begin{equation*}
\{(U(\omega,x,t;m_{\left[0,t\right)}),U^c(\omega,C,t))\}_{\omega\in\Omega,x\in\mathbb{R}_+,C\in\mathbb{R}_+,t\geq 0,m_{\left[0,t\right)}\in\mathcal{P}\left(U_{\sigma}\right)_{\left[0,t\right)}}
\end{equation*}
defined by, for any $x\in\mathbb{R}_+$, $C\in\mathbb{R}_+$, $t\geq 0$, and $m_{\left[0,t\right)}\in\mathcal{P}\left(U_{\sigma}\right)_{\left[0,t\right)}$,
\begin{equation}
U\left(x,t;m_{\left[0,t\right)}\right)=\frac{x^{\kappa}}{\kappa}e^{Y^m_t-g^m_t}\quad\text{and}\quad U^c(C,t)=\frac{C^{\kappa}}{\kappa}\lambda_t,
\label{eq:forward_utility_drift_vol}
\end{equation}
are the robust randomized forward investment and consumption preferences, with drift and volatility uncertainties. Moreover, the optimal and robust forward investment and consumption strategies $\left(\pi^*,c^*\right)\in\mathcal{A}$, in the auxiliary financial market, are given by, for any $t\geq 0$ and $m_{\left[0,t\right)}\in\mathcal{P}\left(U_{\sigma}\right)_{\left[0,t\right)}$,
\begin{equation}
\pi^*_t=x^*_{\pi}\left(t,Z^{m}_t,\bar{Z}^{m}_t\right),\quad c^*_t=\lambda_t^{\frac{1}{1-\kappa}}e^{-\frac{Y^{m}_t}{1-\kappa}}e^{\frac{g^{m}_t}{1-\kappa}},
\label{eq:pi_star}
\end{equation}
where $x^*_{\pi}$ is given in the saddle point of Lemma \ref{lemma:H_saddle_point}.

(ii) The robust randomized preferences in \eqref{eq:forward_utility_drift_vol} satisfies, not only the condition (iii) in Definition \ref{def:forward_performance_drift_vol_random}, but also the following: for any $t\geq 0$, $\xi\in\mathcal{L}\left(\mathcal{F}_t;\mathbb{R}_+\right)$, $m_{\left[0,t\right)}\in\mathcal{P}\left(U_{\sigma}\right)_{\left[0,t\right)}$, and $T\geq t$,
\begin{equation}
\begin{aligned}
&\;U\left(\xi,t;m_{\left[0,t\right)}\right)\\=&\;\esssup_{\left(\pi,c\right)\in\mathcal{A}}\;\essinf_{\left(b,m\right)_{\left[t,T\right)}\in\left(\mathcal{B}\times\mathcal{P}\left(U_{\sigma}\right)\right)_{\left[t,T\right)}}\\&\;\mathbb{E}\left[U\left(X_T^{\xi,t;\pi,c;b,m},T;m_{\left[0,T\right)}\right)+\int_{t}^{T}U^c\left(c_sX_s^{\xi,t;\pi,c;b,m},s\right)ds\vert\mathcal{F}_t\right]\\=&\;\essinf_{\left(b,m\right)_{\left[t,T\right)}\in\left(\mathcal{B}\times\mathcal{P}\left(U_{\sigma}\right)\right)_{\left[t,T\right)}}\\&\;\mathbb{E}\left[U\left(X_{T}^{\xi,t;\pi^*,c^*;b,m},T;m_{\left[0,T\right)}\right)+\int_{t}^{T}U^c(c^*_sX_s^{\xi,t;\pi^*,c^*;b,m},s)ds\vert\mathcal{F}_t\right]\\=&\;\mathbb{E}\left[U\left(X_T^{\xi,t;\pi^*,c^*;b^*,m^*},T;m^*_{\left[0,T\right)}\right)+\int_{t}^{T}U^c(c^*_sX_s^{\xi,t;\pi^*,c^*;b^*,m^*},s)ds\vert\mathcal{F}_t\right]\\=&\; \esssup_{\left(\pi,c\right)\in\mathcal{A}}\mathbb{E}\left[U\left(X_T^{\xi,t;\pi,c;b^*,m^*},T;m^*_{\left[0,T\right)}\right)+\int_{t}^{T}U^c(c_sX_s^{\xi,t;\pi,c;b^*,m^*},s)ds\vert\mathcal{F}_t\right]\\=&\;\essinf_{\left(b,m\right)_{\left[t,T\right)}\in\left(\mathcal{B}\times\mathcal{P}\left(U_{\sigma}\right)\right)_{\left[t,T\right)}}\;\esssup_{\left(\pi,c\right)\in\mathcal{A}}\\&\;\mathbb{E}\left[U\left(X_T^{\xi,t;\pi,c;b,m},T;m_{\left[0,T\right)}\right)+\int_{t}^{T}U^c\left(c_sX_s^{\xi,t;\pi,c;b,m},s\right)ds\vert\mathcal{F}_t\right],
\end{aligned}
\label{eq:summary}
\end{equation}
where $m^*_{\left[0,T\right)}=m_{\left[0,t\right)}\oplus m^*_{\left[t,T\right)}$, for any $s\geq t$,
\begin{equation}
b^*_s=x^*_{b}\left(s,Z^{m^*_{\left[0,s\right)}}_s,\bar{Z}^{m^*_{\left[0,s\right)}}_s\right),\quad m^*_s=x^*_{m}\left(s,Z^{m^*_{\left[0,s\right)}}_s,\bar{Z}^{m^*_{\left[0,s\right)}}_s\right),
\label{eq:worst_b_m}
\end{equation}
$x^*_{b}$ and $x^*_{m}$ are given in the saddle point of Lemma \ref{lemma:H_saddle_point}, and $\left(Z^{m^*_{\left[0,s\right)}}_s,\bar{Z}^{m^*_{\left[0,s\right)}}_s\right)$ depend on $m^*_{\left[0,s\right)}=m_{\left[0,t\right)}\oplus m^*_{\left[t,s\right)}\in\mathcal{P}\left(U_{\sigma}\right)_{\left[0,s\right)}$, which depends on the fixed $t\geq 0$ and $m_{\left[0,t\right)}\in\mathcal{P}\left(U_{\sigma}\right)_{\left[0,t\right)}$.
\end{theorem}

{We offer the following three remarks regarding the theorem above.}

\begin{remark}
Recall that the agent lives in the auxiliary financial market. At the current time $t=0$, the agent first solves a class of infinite-horizon BSDEs and ODEs, parameterized by all possible endogenous randomization processes $m \in \mathcal{P}(U_\sigma)$. Each randomization process for the idiosyncratic volatility in the auxiliary market corresponds to a unique solution $\left(Y^m, Z^m, \bar{Z}^m\right)$ from the infinite-horizon BSDE \eqref{eq:power_infinite_time_horizon_BSDE_drift_vol}, and a unique solution $g^m$ from the ODE \eqref{eq:g_t_nonlinear_ODE_drift_vol}.

The agent's preferences and strategies are based on market-realized randomization, and are designed to adapt to worst-case scenarios. Specifically, the functions $U$ and $U^c$ in \eqref{eq:forward_utility_drift_vol} represent the agent’s robust randomized preferences. These functions depend on the market-realized (instead of the worst-case) randomization from the current time $0$ to (but not including) the future time $t$, as expressed through $Y^m$ and $g^m$.

At time $t$, the agent's optimal and robust investment strategy, given by \eqref{eq:pi_star}, is derived from the saddle point in Lemma \ref{lemma:H_saddle_point}. This saddle point accounts for the worst-case (instead of the market-realized) drift and randomization at time $t$, but depends on the market-realized (instead of the worst-case) randomization from time $0$ to (but not including) time $t$ through $Z^m$ and $\bar{Z}^m$. The optimal and robust consumption strategy at time $t$, as given in \eqref{eq:pi_star}, indirectly depends on the saddle point of the investment strategy and the worst-case drift and randomization. More precisely, it is influenced by the saddle value, which is determined by $Y^m$ and $g^m$.
\end{remark}

\begin{remark}
Note that both the robust randomized preferences and the optimal strategies depend on the correlation coefficient $\rho^i\in\left[-1,1\right]$, for $i=1,\dots,n$. One can interpret it as the agent's sensitivity with respect to the (randomized) idiosyncratic volatility process. Indeed, the optimal investment strategy in \eqref{eq:saddle_point_pi}, with the worst-case drift and randomization, consists of (dropping the projection function and denominator) the myopic component $x_b^{i,*}-r$, the hedging component for the (randomized) idiosyncratic volatility process $z^i\int_{U_{\sigma}^i}ux_m^{i,*}\left(du\right)$, and the hedging component for the systematic volatility process $\bar{\sigma}_t^i\bar{z}$. The second component is weighed by the correlation coefficient $\rho^i$. Note that such dependency on $\rho^i$
exists regardless of the financial market model uncertainty.
\end{remark}

\begin{remark}
The condition on the {process} $\lambda$ in (\ref{eq:ODE_condition_drift_vol}) has a natural interpretation: it provides an implicit upper bound for the forward preference of consumption, determined by the forward preference of investment. However, the choice of $\lambda$ is not unique. For instance, if $\lambda \equiv 0$, or if $\lambda_t$ is {dominated by a} non-increasing {function in time} $t \geq 0$ (thus allowing it to be interpreted as a {stochastic} discounting {factor}), the condition in (\ref{eq:ODE_condition_drift_vol}) is satisfied. See Example \ref{example_lambda} in the next section for further details.
\end{remark}

\begin{example} (Continuation of Example \ref{example})
\label{eg:section_4} In the auxiliary financial market, the stochastic factor model for $V^m=(V^{1,m},\dots,V^{n,m})$ as a result of the randomization by $m\in\mathcal{P}\left(U_{\sigma}\right)$ is given by, for each $i=1,\dots,n$, and for any $t\geq 0$,
$$dV_t^{i,m}=\eta^i(V_t^m)dt+\kappa^i(V_t^m)dB_t^{i,m^i}+\bar{\kappa}^i(V_t^m)d\bar{W}_t.$$
The solution $(Y^m, Z^m,\bar{Z}^m)$ of BSDE (\ref{eq:power_infinite_time_horizon_BSDE_drift_vol}) can then be represented via some deterministic functions due to the Markov property of the model. Indeed, there exists a measurable function $y(\cdot)$ such that $Y^m_t=y(V_t^m)$, for $t\geq 0$. Suppose that $y(\cdot)$ is twice differentiable, with $\partial_{v^i}y(\cdot)$ as the first partial derivative with respect to $v^i$, for $i=1,\dots,n$, and $\partial_{v^iv^j}y(\cdot)$ as the second partial derivative with respect to $v^i$ and $v^j$, for $i,j=1,\dots,n$. By It\^o's formula, we have
\begin{align*}
d y(V_t^m)=&\; \frac12\sum_{i=1}^n\left(\kappa^i(V_t^m)^2+\bar{\kappa}^i(V_t^m)^2\right)\partial_{v^iv^i}y(V_t^m)dt\\
&+\sum_{i=1}^{n}\sum_{j=i+1}^{n}\bar{\kappa}^i(V_t^m)\bar{\kappa}^j(V_t^m)\partial_{v^iv^j}y(V_t^m)dt+
\sum_{i=1}^n\eta^i(V_t^m)\partial_{v^i}y(V_t^m)dt\\
&+\sum_{i=1}^n\partial_{v^i}y(V_t^m)\left(\kappa^i(V_t^m)dB_t^{i,m^i}+\bar{\kappa}^i(V_t^m)d\bar{W}_t\right).
\end{align*}
Identifying the martingale and the finite variation parts of $y(V^m)$ with the BSDE (\ref{eq:power_infinite_time_horizon_BSDE_drift_vol}), we further obtain that, for $t\geq 0$,
$$Z_t^{i,m}=\partial_{v^i}y(V_t^m)\kappa^i(V_t^m),\ i=1,\dots,n,$$
and $\bar{Z}_t^m=\sum_{i=1}^n\partial_{v^i}y(V_t^m)\bar{\kappa}^i(V_t^m)$. In turn, we derive the following elliptic PDE for the characterization of the robust randomized forward preferences: for any $v\in\mathbb{R}^n$,
\begin{align}\label{PDE}
&\frac12\sum_{i=1}^n\left(\kappa^i(v)^2+\bar{\kappa}^i(v)^2\right)\partial_{v^iv^i}y(v)+\sum_{i=1}^{n}\sum_{j=i+1}^{n}\bar{\kappa}^i(v)\bar{\kappa}^j(v)\partial_{v^iv^j}y(v)+\sum_{i=1}^n\eta^i(v)\partial_{v^i}y(v)\notag\\
&-\rho y(v)+H^*\left(v,\left(\partial_{v^1}y(v){\kappa}^1(v),\dots,\partial_{v^n}y(v){\kappa}^n(v)\right),\sum_{i=1}^n\partial_{v^i}y(v)\bar{\kappa}^i(v)\right)=0.
\end{align}
{The robust randomized forward investment preference and the corresponding optimal and robust investment strategy are then given by
\begin{equation*}
U(x, t; m_{[0,t)}) = \frac{x^{\kappa}}{\kappa} e^{y(V^m_t) - g^m_t},
\end{equation*}
\begin{equation*}
\pi^*_t = x^*_{\pi} \left( t, \left(\partial_{v^1} y(V^m_t) \kappa^1(V^m_t),\dots,\partial_{v^n} y(V^m_t) \kappa^n(V^m_t)\right), \sum_{i=1}^n \partial_{v^i} y(V^m_t) \bar{\kappa}^i(V^m_t) \right),
\end{equation*}
for all \( x \in \mathbb{R}_+ \), \( t \geq 0 \), and \( m_{[0,t)} \in \mathcal{P}(U_\sigma)_{[0,t)} \).}

\end{example}

\section{Robust CRRA Forward Preferences}\label{sec:5}
Armed with the constructed robust randomized forward preferences and the corresponding optimal and robust strategies in the auxiliary financial market, we now proceed to demonstrate that these preferences and strategies are also robust forward preferences and optimal and robust strategies for the agent in the physical financial market. In particular, we will begin by proving the following proposition, which further strengthens \eqref{eq:summary} from Theorem \ref{thm:power_forward_utility_consumption_drift_vol}.

\begin{proposition}\label{new_prop}
Assume that the conditions in Theorem \ref{thm:power_forward_utility_consumption_drift_vol} hold, and let $\left(Y^m,Z^m,\bar{Z}^m\right)$ and $g^m$ be the solutions of the infinite-horizon BSDE \eqref{eq:power_infinite_time_horizon_BSDE_drift_vol} and the ODE \eqref{eq:g_t_nonlinear_ODE_drift_vol} respectively. The robust randomized forward investment and consumption preferences in \eqref{eq:forward_utility_drift_vol} satisfies that, for any $t\geq 0$, $\xi\in\mathcal{L}\left(\mathcal{F}_t;\mathbb{R}_+\right)$, $m_{\left[0,t\right)}\in\mathcal{P}\left(U_{\sigma}\right)_{\left[0,t\right)}$, and $T\geq t$,
\begin{equation}
\begin{aligned}
&\;U\left(\xi,t;m_{\left[0,t\right)}\right)\\=&\;\essinf_{\left(b,m\right)_{\left[t,T\right)}\in\left(\mathcal{B}\times\mathcal{P}'\left(U_{\sigma}\right)\right)_{\left[t,T\right)}}\\&\;\mathbb{E}\left[U\left(X_{T}^{\xi,t;\pi^*,c^*;b,m},T;m_{\left[0,T\right)}\right)+\int_{t}^{T}U^c(c^*_sX_s^{\xi,t;\pi^*,c^*;b,m},s)ds\vert\mathcal{F}_t\right]\\=&\;\esssup_{\left(\pi,c\right)\in\mathcal{A}}\;\essinf_{\left(b,m\right)_{\left[t,T\right)}\in\left(\mathcal{B}\times\mathcal{P}'\left(U_{\sigma}\right)\right)_{\left[t,T\right)}}\\&\;\mathbb{E}\left[U\left(X_T^{\xi,t;\pi,c;b,m},T;m_{\left[0,T\right)}\right)+\int_{t}^{T}U^c\left(c_sX_s^{\xi,t;\pi,c;b,m},s\right)ds\vert\mathcal{F}_t\right],
\end{aligned}
\label{eq:summary_expand}
\end{equation}
where, in the second line of \eqref{eq:summary_expand}, $\left(\pi^*,c^*\right)_{\left[t,T\right)}\in\mathcal{A}_{\left[t,T\right)}$ is given by \eqref{eq:pi_star}, in which, for any $s\in\left[t,T\right)$, $\left(\pi^*_s,c^*_s\right)$ depends on $m_{\left[0,s\right)}=m_{\left[0,t\right)}\oplus m_{\left[t,s\right)}$, with $m_{\left[0,t\right)}\in\mathcal{P}\left(U_{\sigma}\right)_{\left[0,t\right)}$ and $m_{\left[t,s\right)}\in\mathcal{P}'\left(U_{\sigma}\right)_{\left[t,s\right)}$.
\end{proposition}

Recall that, when $m\in\mathcal{P}'\left(U_{\sigma}\right)$, the financial market is the physical one. Therefore, Theorem \ref{thm:power_forward_utility_consumption_drift_vol} and Proposition \ref{new_prop} together immediately imply the following main theorem in this paper, which constructs the robust forward preferences, and optimal and robust strategies of the agent in the physical financial market.

\begin{theorem}\label{thm:power_forward_utility_consumption_drift}
For any $\rho>0$, let $\left(Y,Z,\bar{Z}\right)$ be the solution of the infinite-horizon BSDE that, for any $t\geq 0$,
\begin{equation}
dY_t=-\left(H^*\left(t,Z_t,\bar{Z}_t\right)-\rho Y_t\right)dt+\sum_{i=1}^{n}Z_t^{i}dB_t^i+\bar{Z}_td\bar{W}_t;
\label{eq:power_infinite_time_horizon_BSDE_drift}
\end{equation}
let $g_t$, $t\geq 0$, be the solution of the following ODE: for any $t\geq 0$,
\begin{equation}
dg_t=\left(\left(1-\kappa\right)\lambda_t^{\frac{1}{1-\kappa}}e^{-\frac{Y_t}{1-\kappa}}e^{\frac{g_t}{1-\kappa}}+\rho Y_t\right)dt,
\label{eq:condition_1}
\end{equation}
where the {uniformly bounded and $\mathbb{F}$-progressively measurable process} 
$\lambda_t$, $t\geq 0$, satisfies the condition that, for any $t\geq 0$,
\begin{equation}
e^{-\frac{g_0}{1-\kappa}}>\int_{0}^{t}e^{\frac{1}{1-\kappa}\left(\rho\int_{0}^{s}Y_udu-Y_s\right)}\lambda_s^{\frac{1}{1-\kappa}}ds.
\label{eq:condition_2}
\end{equation}
Then the pair of processes
\begin{equation*}
\{(U(\omega,x,t),U^c(\omega,C,t))\}_{\omega\in\Omega,x\in\mathbb{R}_+,C\in\mathbb{R}_+,t\geq 0}
\end{equation*}
defined by, for any $x\in\mathbb{R}_+$, $C\in\mathbb{R}_+$, and $t\geq 0$,
\begin{equation*}
U\left(x,t\right)=\frac{x^{\kappa}}{\kappa}e^{Y_t-g_t}\quad\text{and}\quad U^c(C,t)=\frac{C^{\kappa}}{\kappa}\lambda_t,
\end{equation*}
are the robust forward investment and consumption preferences, with drift and volatility uncertainties. Moreover, the optimal and robust forward investment and consumption strategies $\left(\pi^*,c^*\right)\in\mathcal{A}$, in the physical financial market, are given by, for any $t\geq 0$,
\begin{equation*}
\pi^*_t=x^*_{\pi}\left(t,Z_t,\bar{Z}_t\right),\quad c^*_t=\lambda_t^{\frac{1}{1-\kappa}}e^{-\frac{Y_t}{1-\kappa}}e^{\frac{g_t}{1-\kappa}},
\end{equation*}
where $x^*_{\pi}$ is given in the saddle point of Lemma \ref{lemma:H_saddle_point}.
\end{theorem}

{We offer the following two remarks regarding the theorem above.

\begin{remark}
The financial interpretation of this main result is similar to that in Section \ref{sec:4}, where the agent lives in the auxiliary financial market. Herein, the agent lives in the physical market.

At time $t=0$, the agent only needs to solve one infinite-horizon BSDE \eqref{eq:power_infinite_time_horizon_BSDE_drift} and one ODE \eqref{eq:condition_1} without randomization. Her robust preferences and the optimal strategies depend on the market-realized past idiosyncratic volatility, instead of its randomization, as well as the current saddle-value given in Lemma \ref{lemma:H_saddle_point}. Similar to an agent living in the auxiliary market, at each future time, the agent living in the physical market optimally invests according to the saddle point in Lemma \ref{lemma:H_saddle_point}. Specifically, her best investment strategy accounts for the worst-case drift and the worst-case randomization on the idiosyncratic volatility, rather than for the volatility itself.

Although the agent's optimal and robust consumption strategy does not depend on randomization anymore, it still depends on the current saddle-value through $Y$ and $g$ by the BSDE \eqref{eq:power_infinite_time_horizon_BSDE_drift} and the ODE \eqref{eq:condition_1}. Her belief about the consumption preferences via $\lambda$ affects both her consumption strategy as well as her investment preferences. Her optimal and robust consumption strategy is proportional to this belief. When the agent weighs more on her consumption preferences (i.e., when the agent discounts less her utility gain through consumption, if $\lambda$ is interpreted as a stochastic discounting factor as in Example \ref{example_lambda} below), the utility gain through her investment preferences reduces, assuming that the same process $Y$ still satisfies the structural condition \eqref{eq:condition_2}. This can be seen clearly by \eqref{eq:solution_gg} in the theorem's proof.
\end{remark}}

\begin{remark}
It is easy to observe from the BSDE \eqref{eq:power_infinite_time_horizon_BSDE_drift} that, even if $U_b$ and $U_{\sigma}$ are singletons, the robust preferences and the optimal strategies still depend on the correlation coefficient $\rho^i$, for $i=1,\dots,n$. This echoes that the agent's sensitivity with respect to the idiosyncratic volatility process exists regardless of her ambiguity aversion on the financial market model.
\end{remark}

\begin{example} (Continuation of Example \ref{example}) In the stochastic factor model, following along the similar arguments in Example \ref{eg:section_4}, one can obtain a Markovian representation for the solution of the BSDE (\ref{eq:power_infinite_time_horizon_BSDE_drift}). That is, $Y_t=y(V_t)$, for $t\geq 0$, with $y(\cdot)$ satisfying the elliptic PDE (\ref{PDE}), under the assumption that the solution function $y(\cdot)$ is twice differentiable.
\end{example}

\begin{example}\label{example_lambda}
({Stochastic} discounting {factor} $\lambda$) We provide a rich class of {uniformly bounded and $\mathbb{F}$-progressively measurable processes} $\lambda$ which satisfy the condition \eqref{eq:condition_2}. Suppose that the {process} $\lambda$ is given by, for any $t\geq 0$,
\begin{equation*}
{\lambda_t=\alpha_te^{-\left(\rho D_t+\beta_t\right)t},}
\end{equation*}
{for some $\mathbb{F}$-progressively measurable processes $\alpha_t,\beta_t,D_t$, for $t\geq 0$. Assume that $\alpha$ is non-negative and uniformly upper-bounded by a constant $\overline{\alpha}\geq 0$; $\beta$ is uniformly lower-bounded by another constant $\underline{\beta}>0$; and $D$ is uniformly lower-bounded by the constant $\underline{D}>0$, which}  
is the uniformly bounded constant for the solution $Y$ of the infinite-horizon BSDE \eqref{eq:power_infinite_time_horizon_BSDE_drift}. Then, sufficiently, condition (\ref{eq:condition_2}) is satisfied when
\begin{equation*}
e^{-\frac{g_0}{1-\kappa}}>\frac{1-\kappa}{\underline{\beta}}{\overline{\alpha}}^{\frac{1}{1-\kappa}}e^{\frac{\underline{D}}{1-\kappa}}.
\end{equation*}
\end{example}

{Finally, the following corollary of Theorem~\ref{thm:power_forward_utility_consumption_drift} recovers the (robust) forward preferences and associated optimal strategies for three well-studied special cases in the literature:

(i) Forward investment preference \cite{MZ0,MZ-Kurtz,MZ-Carmona,MZ1,MZ2,MZ3}:
    Obtained when $U^c \equiv 0$ and $\mathcal{B} \times \Sigma$ is a singleton.

(ii) Forward investment and consumption preference \cite{Berrier_2008,El_Karoui_2018,El_Karoui_2024,Kallblad_2016}: 
    Obtained when  $U^c \not\equiv 0$ with $\mathcal{B} \times \Sigma$ a singleton.

(iii) Robust forward investment preference \cite{Kallblad_2018,Lin_2020}: Obtained when  $U^c \equiv 0$ and $\mathcal{B} \times \Sigma$ is not a singleton.}

\begin{corollary}
For any $\rho>0$, let $\left(Y',Z',\bar{Z}'\right)$ be the solution of the infinite-horizon BSDE that, for any $t\geq 0$,
\begin{equation}\label{eq:power_infinite_time_horizon_BSDE}
dY_t=-\left(\left(H^{*}\right)'\left(t,Z_t,\bar{Z}_t\right)-\rho Y_t\right)dt+\sum_{i=1}^{n}Z_t^{i}dB_t^i+\bar{Z}_td\bar{W}_t,
\end{equation}
where $\left(H^{*}\right)'$ is the maximum value of the Hamiltonian given in (\ref{eq:H_canonical}), for some fixed $\left(b,\sigma\right)\in\mathcal{B}\times\Sigma$, that is, for any $\left(\omega,t,z,\bar{z}\right)\in\Omega\times[0,\infty)\times\mathbb{R}^n\times\mathbb{R}$,
$$\left(H^{*}\right)'(\omega,t,z,\bar{z})=\sup_{x_{\pi}\in\Pi}H(\omega, t,z,\bar{z};x_{\pi};b_t,\sigma_t),$$
which is given as in \eqref{eq:H_sub_optimal} by replacing $x^i_b$ by $b^i_t$, and the integrals with $\sigma^i_t$ and $\left(\sigma^i_t\right)^2$, for $i=1,\dots,n$. Moreover, let $(Y,Z,\bar{Z})$ be the solution of the BSDE (\ref{eq:power_infinite_time_horizon_BSDE_drift}). The following assertions hold.

(i) Suppose that $U^c\equiv 0$ and $\mathcal{B}\times\Sigma$ is a singleton. Then, the process $U\left(x,t\right)=\frac{x^{\kappa}}{\kappa}e^{Y_t'-\rho\int_{0}^{t}Y'_sds}$, $x\in\mathbb{R}_+$, $t\geq 0$, is a forward investment preference. Moreover, the optimal forward investment strategy is given by, for any $t\geq 0$,
\begin{equation*}
\pi^*_t=\left(x^{*}_{\pi}\right)'\left(t,Z_t',\bar{Z}_t'\right),
\end{equation*}
where, for each $i=1,\dots,n$, $\left(x^{i,*}_{\pi}\right)'$ is given as, for any $t\geq 0$, $z^i\in\mathbb{R}$, and $\bar{z}\in\mathbb{R}$,
\begin{equation}
\left(x^{i,*}_{\pi}\right)'\left(t,z^i,\bar{z}\right)=\text{Proj}_{\Pi^i}\left(\frac{b^i_t-r+\rho^iz^i\sigma^i_t+\bar{\sigma}_t^i\bar{z}}{\left(1-\kappa\right)\left(\left(\sigma^i_t\right)^2+\left(\bar{\sigma}_t^i\right)^2\right)}\right),
\label{eq:x^*'}
\end{equation}
which is the same as in \eqref{eq:saddle_point_pi} by replacing $x^i_b$ by $b^i_t$, and the integrals with $\sigma^i_t$ and $\left(\sigma^i_t\right)^2$, for $i=1,\dots,n$; these coincide with Theorem 3.6 of \cite{LZ} in the stochastic factor model.

(ii) Suppose that $U^{c}\not\equiv 0$ and  $\mathcal{B}\times\Sigma$ is a singleton. Then, the pair of processes
\begin{equation*}
\{(U(\omega,x,t),U^c(\omega,C,t))\}_{\omega\in\Omega,x\in\mathbb{R}_+,C\in\mathbb{R}_+,t\geq 0}
\end{equation*}
defined by, for any $x\in\mathbb{R}_+$, $C\in\mathbb{R}_+$, and $t\geq 0$,
\begin{equation*}
U\left(x,t\right)=\frac{x^{\kappa}}{\kappa}e^{Y_t'-g_t'}\quad\text{and}\quad U^c(C,t)=\frac{C^{\kappa}}{\kappa}\lambda_t',
\end{equation*}
are forward investment and consumption preferences. Moreover, the optimal forward investment and consumption strategies $\left(\pi^*,c^*\right)$ are given by, for any $t\geq 0$,
\begin{equation*}
\pi^*_t=\left(x^{*}_{\pi}\right)'\left(t,Z_t',\bar{Z}_t'\right),\quad c^*_t=(\lambda_t')^{\frac{1}{1-\kappa}}e^{-\frac{Y_t'}{1-\kappa}}e^{\frac{g_t'}{1-\kappa}},
\end{equation*}
where $\left(x^{*}_{\pi}\right)'$ is given in \eqref{eq:x^*'}, $\lambda'$ satisfies the condition (\ref{eq:condition_2}) with $Y'$, and $g'$ is the solution of the ODE \eqref{eq:condition_1} with $\lambda'$ and $Y'$.

(iii) Suppose that $U^c\equiv 0$ and $\mathcal{B}\times\Sigma$ is not a singleton. Then, the process $U\left(x,t\right)=\frac{x^{\kappa}}{\kappa}e^{Y_t-\rho\int_{0}^{t}Y_sds}$, $x\in\mathbb{R}_+$, $t\geq 0$, is a robust forward investment preference. Moreover, the optimal and robust forward investment strategy is given by, for any $t\geq 0$,
\begin{equation*}
\pi^*_t=x^{*}_{\pi}\left(t,Z_t,\bar{Z}_t\right),
\end{equation*}
where $x^{*}_{\pi}$ is given in the saddle point of Lemma
\ref{lemma:H_saddle_point}.
\end{corollary}


{
\renewcommand*\appendixpagename{\Large Appendices}
\section{Proofs}}

{\subsection{Existence of Saddle Points: Proof of Lemma \ref{lemma:H_saddle_point}}}
Fix any $\omega\in\Omega$, $t\geq 0$, $z\in\mathbb{R}^n$, $\bar{z}\in\mathbb{R}$. For any $\left(x_{b},x_{m}\right)\in U_b\times\mathcal{P}\left(U_{\sigma}\right)_{\omega,t}$, $H\left(\omega,t,z,\bar{z};\cdot;x_{b},x_{m}\right)$ in \eqref{eq:H} is concave and continuous in $x_{\pi}\in\Pi$, where $\Pi$ is convex and closed; for any $x_{\pi}\in\Pi$, $H\left(\omega,t,z,\bar{z};x_{\pi};\cdot,\cdot\right)$ is linear and continuous in $\left(x_{b},x_{m}\right)\in U_b\times\mathcal{P}\left(U_{\sigma}\right)_{\omega,t}$, where $U_b$ and $\mathcal{P}\left(U_{\sigma}\right)_{\omega,t}$ are convex and compact. By, for example, Theorem 2.132 in \cite{Barbu_2012}, (i) and (ii) follow. For any $\left(x_{b},x_{m}\right)\in U_b\times\mathcal{P}\left(U_{\sigma}\right)_{\omega,t}$, the maximization problem $\sup_{x_{\pi}\in\Pi}H\left(\omega,t,z,\bar{z};x_{\pi};x_{b},x_{m}\right)$ easily yields \eqref{eq:saddle_point_pi}. Since $H\left(\omega,t,z,\bar{z};x_{\pi}^{*}\left(x_{b},x_{m}\right);x_{b},x_{m}\right)$ in \eqref{eq:H_sub_optimal} is continuous in $\left(x_{b},x_{m}\right)\in U_b\times\mathcal{P}\left(U_{\sigma}\right)_{\omega,t}$ (as the distance function is $1$-Lipschitz continuous in $a\in\mathbb{R}$), where $U_b$ and $\mathcal{P}\left(U_{\sigma}\right)_{\omega,t}$ are convex and compact, $\left(x_{b}^{*},x_{m}^{*}\right)$ in \eqref{eq:saddle_point_b_m} exists.

For the saddle value function $H^*\left(\cdot,\cdot,\cdot,\cdot\right)$, the first and second equalities in \eqref{eq:saddle_valeu_equivalence} are the definition because of (i), the third equality is due to the existence of a saddle point in (ii), while the second last and last equalities are based on the explicit saddle-point in (iii).\\

{\subsection{Solvability of  Randomized Infinite-Horizon BSDE: Proof of Proposition \ref{prop:power_infinite_BSDE_existence_drift_vol}}}
By Lemma \ref{lemma:H_saddle_point}, the saddle value function $H^*\left(\cdot,\cdot,\cdot,\cdot\right)$ is given by, for any $\left(\omega,t,z,\bar{z}\right)\in\Omega\times[0,\infty)\times\mathbb{R}^n\times\mathbb{R}$,
\begin{equation*}
H^*\left(\omega,t,z,\bar{z}\right)=\inf_{\left(x_{b},x_{m}\right)\in U_b\times\mathcal{P}\left(U_{\sigma}\right)_{\omega,t}}H\left(\omega,t,z,\bar{z};x_{\pi}^{*}\left(x_{b},x_{m}\right);x_{b},x_{m}\right),
\end{equation*}
where $H\left(\omega,t,z,\bar{z};x_{\pi}^{*}\left(x_{b},x_{m}\right);x_{b},x_{m}\right)$, for any $\left(x_{b},x_{m}\right)\in U_b\times\mathcal{P}\left(U_{\sigma}\right)_{\omega,t}$, is given by \eqref{eq:H_sub_optimal}. Hence, for any $\left(\omega,t\right)\in\Omega\times[0,\infty)$, $\left(z_1,\bar{z}_1\right)\in\mathbb{R}^{n+1}$, and $\left(z_2,\bar{z}_2\right)\in\mathbb{R}^{n+1}$, with the respective $\left(x_{b,1}^{*},x_{m,1}^{*}\right)\in U_b\times\mathcal{P}\left(U_{\sigma}\right)_{\omega,t}$ and $\left(x_{b,2}^{*},x_{m,2}^{*}\right)\in U_b\times\mathcal{P}\left(U_{\sigma}\right)_{\omega,t}$ defined in \eqref{eq:saddle_point_b_m},
\begin{align*}
&\;H^*\left(\omega,t,z_1,\bar{z}_1\right)-H^*\left(\omega,t,z_2,\bar{z}_2\right)\\=&\;\inf_{\left(x_{b},x_{m}\right)\in U_b\times\mathcal{P}\left(U_{\sigma}\right)_{\omega,t}}H\left(\omega,t,z_1,\bar{z}_1;x_{\pi}^{*}\left(x_{b},x_{m}\right);x_{b},x_{m}\right)\\&\;-\inf_{\left(x_{b},x_{m}\right)\in U_b\times\mathcal{P}\left(U_{\sigma}\right)_{\omega,t}}H\left(\omega,t,z_2,\bar{z}_2;x_{\pi}^{*}\left(x_{b},x_{m}\right);x_{b},x_{m}\right)\\\leq&\;H\left(\omega,t,z_1,\bar{z}_1;x_{\pi}^{*}\left(x_{b,2}^*,x_{m,2}^*\right);x_{b,2}^*,x_{m,2}^*\right)\\&\;-H\left(\omega,t,z_2,\bar{z}_2;x_{\pi}^{*}\left(x_{b,2}^*,x_{m,2}^*\right);x_{b,2}^*,x_{m,2}^*\right);
\end{align*}
and thus, by \eqref{eq:H_sub_optimal},
\begin{align*}
&\;H^*\left(\omega,t,z_1,\bar{z}_1\right)-H^*\left(\omega,t,z_2,\bar{z}_2\right)\\\leq&\;\sum_{i=1}^{n}\left(-\frac{1}{2}\kappa\left(1-\kappa\right)\left(\int_{U_{\sigma}^{i}}u^2x_{m,2}^{i,*}\left(du\right)+\left(\bar{\sigma}_t^i\right)^2\right)\right.\\&\;\quad\quad\quad\quad\left.\times\left(\text{dist}^2\left\{\Pi^i,\frac{x_{b,2}^{i,*}-r+\rho^iz_1^i\int_{U_{\sigma}^i}ux_{m,2}^{i,*}\left(du\right)+\bar{\sigma}_t^i\bar{z}_1}{\left(1-\kappa\right)\left(\int_{U_{\sigma}^i}u^2x_{m,2}^{i,*}\left(du\right)+\left(\bar{\sigma}_t^i\right)^2\right)}\right\}\right.\right.\\&\;\left.\left.\quad\quad\quad\quad\quad\quad-\;\text{dist}^2\left\{\Pi^i,\frac{x_{b,2}^{i,*}-r+\rho^iz_2^i\int_{U_{\sigma}^i}ux_{m,2}^{i,*}\left(du\right)+\bar{\sigma}_t^i\bar{z}_2}{\left(1-\kappa\right)\left(\int_{U_{\sigma}^i}u^2x_{m,2}^{i,*}\left(du\right)+\left(\bar{\sigma}_t^i\right)^2\right)}\right\}\right)\right.\\&\;\quad\quad\;\left.+\frac{1}{2}\frac{\kappa}{1-\kappa}\frac{1}{\int_{U_{\sigma}^i}u^2x_{m,2}^{i,*}\left(du\right)+\left(\bar{\sigma}_t^i\right)^2}\right.\\&\;\quad\quad\quad\quad\left.\times\left(\left(x_{b,2}^{i,*}-r+\rho^iz_1^i\int_{U_{\sigma}^i}ux_{m,2}^{i,*}\left(du\right)+\bar{\sigma}_t^i\bar{z}_1\right)^2\right.\right.\\&\;\left.\left.\quad\quad\quad\quad\quad\quad-\left(x_{b,2}^{i,*}-r+\rho^iz_2^i\int_{U_{\sigma}^i}ux_{m,2}^{i,*}\left(du\right)+\bar{\sigma}_t^i\bar{z}_2\right)^2\right)\right)\\&\;+\frac{1}{2}\sum_{i=1}^{n}\left(\left(z_1^i\right)^2-\left(z_2^i\right)^2\right)+\frac{1}{2}\left(\left(\bar{z}_1\right)^2-\left(\bar{z}_2\right)^2\right),
\end{align*}
with a similar estimate for the lower bound.

By the above estimates, the uniform boundedness of $x_{b,1}^{*},x_{m,1}^{*},x_{b,2}^{*},x_{m,2}^{*},\bar{\sigma}$, the property that $\Pi$ is a convex and compact subset in $\mathbb{R}^n$ including the origin $0\in\mathbb{R}^n$, and the $1$-Lipschitz continuity of the distance function, there exists a constant $K>0$ such that, for any $\left(\omega,t\right)\in\Omega\times[0,\infty)$, $y\in\mathbb{R}$, $\left(z_1,\bar{z}_1\right)\in\mathbb{R}^{n+1}$, and $\left(z_2,\bar{z}_2\right)\in\mathbb{R}^{n+1}$,
\begin{align*}
\vert F\left(t,y,z_1,\bar{z}_1\right)-F\left(t,y,z_2,\bar{z}_2\right)\vert\leq&\; K\left(1+\vert\left(z_1,\bar{z}_1\right)\vert+\vert\left(z_2,\bar{z}_2\right)\vert\right)\\&\;\times\vert \left(z_1,\bar{z}_1\right)-\left(z_2,\bar{z}_2\right)\vert
\end{align*}
and $\vert F(t,0,0,0)\vert\leq K$, where $F$ is the driver of the infinite-horizon BSDE \eqref{eq:power_infinite_time_horizon_BSDE_drift_vol}. Moreover, the driver $F$ is monotone in $y$, in the sense that, for any $\left(\omega,t\right)\in\Omega\times[0,\infty)$, $y_1,y_2\in\mathbb{R}$, $z\in\mathbb{R}^n$, and $\bar{z}\in\mathbb{R}$,
\begin{equation*}
\left(y_1-y_2\right)\left(F\left(t,y_1,z,\bar{z}\right)-F\left(t,y_2,z,\bar{z}\right)\right)\leq-\rho\left(y_1-y_2\right)^2.
\end{equation*}
Therefore, the driver $F$ satisfies Assumption A1 in \cite{Briand_2008}, and hence, we conclude by Theorem 3.3 in \cite{Briand_2008}.\\

{\subsection{Verification for Robust Randomized Forward Preference: Proof of Theorem \ref{thm:power_forward_utility_consumption_drift_vol}}}
With the condition \eqref{eq:ODE_condition_drift_vol}, the ODE \eqref{eq:g_t_nonlinear_ODE_drift_vol} can be uniquely solved. Indeed, by an exponential transformation that, for any $t\geq 0$, $\bar{g}_t=e^{-\frac{g_t}{1-\kappa}}$, the solution $g^m_t$, $t\geq 0$, of the ODE \eqref{eq:g_t_nonlinear_ODE_drift_vol} is uniquely given by, for any $t\geq 0$,
\begin{equation}
g^m_t=\rho\int_{0}^{t}Y^m_sds-\left(1-\kappa\right)\ln\left(e^{\frac{-g^m_0}{1-\kappa}}-\int_{0}^{t}e^{\frac{1}{1-\kappa}\left(\rho\int_{0}^{s}Y^m_udu-Y^m_s\right)}\lambda_s^{\frac{1}{1-\kappa}}ds\right),
\label{eq:solution_gg}
\end{equation}
which, due to the continuity and the uniform boundedness of $Y^m${, as well as the uniform boundedness of $\lambda$}, is locally uniformly bounded in any $\left[0,t\right]$, $t\geq 0$.

We only have to check the condition (iii) in Definition \ref{def:forward_performance_drift_vol_random} as the conditions (i) and (ii) are obviously true. By the facts that $Y^m$ is uniformly bounded and $g^m$ is locally uniformly bounded{, as well as the uniform boundedness of $\lambda$}, $\left(\pi^*,c^*\right)\in\mathcal{A}$ given in \eqref{eq:pi_star}. Clearly, $\left(b^*,m^*\right)\in\mathcal{B}\times\mathcal{P}\left(U_{\sigma}\right)$ given in \eqref{eq:worst_b_m}.

Fix $t\geq 0$, $\xi\in\mathcal{L}\left(\mathcal{F}_t;\mathbb{R}_+\right)$, $m_{\left[0,t\right)}\in\mathcal{P}\left(U_{\sigma}\right)_{\left[0,t\right)}$, and $T\geq t$ in the remaining of this proof. For any $\left(\pi,c\right)_{\left[t,T\right)}\in\mathcal{A}_{\left[t,T\right)}$, and $\left(b,m\right)_{\left[t,T\right)}\in\left(\mathcal{B}\times\mathcal{P}\left(U_{\sigma}\right)\right)_{\left[t,T\right)}$, define
\begin{equation*}
R_T^{\xi,t,m_{\left[0,t\right)};\pi,c;b,m}=U\left(X_T^{\xi,t;\pi,c;b,m},T;m_{\left[0,T\right)}\right)+\int_{t}^{T}U^c(c_sX_s^{\xi,t;\pi,c;b,m},s)ds,
\end{equation*}
where $X^{\xi,t;\pi,c;b,m}$ solves \eqref{wealth_random_2} with $X^{\xi,t;\pi,c;b,m}_t=\xi$, $\left(U,U^c\right)$ are given in \eqref{eq:forward_utility_drift_vol}, and $m_{\left[0,T\right)}=m_{\left[0,t\right)}\oplus m_{\left[t,T\right)}$. In particular, $R_t^{\xi,t,m_{\left[0,t\right)};\pi,c;b,m}=U\left(\xi,t;m_{\left[0,t\right)}\right)$. By the It{\^o}'s formula, $R^{\xi,t,m_{\left[0,t\right)};\pi,c;b,m}$ solves, for any $s\geq t$,
\begin{equation}
\begin{aligned}
dR_s=&\;\frac{X_s^{\kappa}}{\kappa}e^{Y_s-g_s}\Bigg(\bigg(H\left(s,Z_s,\bar{Z}_s;\pi_s;b_s,m_s\right)-H^*\left(s,Z_s,\bar{Z}_s\right)\\&+c_s^{\kappa}e^{-Y_s+g_s}\lambda_s-\kappa c_s-\left(1-\kappa\right)\lambda_s^{\frac{1}{1-\kappa}}e^{-\frac{Y_s}{1-\kappa}}e^{\frac{g_s}{1-\kappa}}\bigg)ds\\&+\sum_{i=1}^{n}\left(\kappa\pi^i_s\sqrt{\int_{U_{\sigma}^i}u^2m_{s}^{i}\left(du\right)}+Z^{i}_s\rho_s^{i,m}\right)dW^i_s\\&+\sum_{i=1}^{n}Z^{i}_s\sqrt{1-\left(\rho_s^{i,m}\right)^2}dW^{i+n}_s+\left(\kappa\sum_{i=1}^{n}\pi^i_s\bar{\sigma}^i_s+\bar{Z}_s\right)d\bar{W}_s\Bigg),
\end{aligned}
\label{eq:R}
\end{equation}
where $H$ and $H^*$ are defined in \eqref{eq:H} and \eqref{eq:saddle_valeu_equivalence}, and we have used the definition of the randomized Brownian motion $B^m$ introduced in (\ref{new_BM}).

On one hand, for any $\left(\pi,c\right)_{\left[t,T\right)}\in\mathcal{A}_{\left[t,T\right)}$ and $s\in\left[t,T\right)$, by Lemma \ref{lemma:H_saddle_point},
\begin{align*}
&\;H\left(s,Z^{m^*}_s,\bar{Z}^{m^*}_s;\pi_s;b^*_s,m^*_s\right)-H^*\left(s,Z^{m^*}_s,\bar{Z}^{m^*}_s\right)\\&+c_s^{\kappa}e^{-Y^{m^*}_s+g^{m^*}_s}\lambda_s-\kappa c_s-\left(1-\kappa\right)\lambda_s^{\frac{1}{1-\kappa}}e^{-\frac{Y^{m^*}_s}{1-\kappa}}e^{\frac{g^{m^*}_s}{1-\kappa}}\leq 0,
\end{align*}
where $\left(b^*,m^*\right)_{\left[t,T\right)}\in\left(\mathcal{B}\times\mathcal{P}\left(U_{\sigma}\right)\right)_{\left[t,T\right)}$ is given in \eqref{eq:worst_b_m}. Therefore, for any $\left(\pi,c\right)_{\left[t,T\right)}\in\mathcal{A}_{\left[t,T\right)}$, $R^{\xi,t,m_{\left[0,t\right)};\pi,c;b^*,m^*}$ is a local $\mathbb{F}$-supermartingale. Since $R^{\xi,t,m_{\left[0,t\right)};\pi,c;b^*,m^*}$ is non-negative, it is even a proper $\mathbb{F}$-supermartingale. Hence, for any $\left(\pi,c\right)_{\left[t,T\right)}\in\mathcal{A}_{\left[t,T\right)}$,
\begin{equation*}
\mathbb{E}\left[R_T^{\xi,t,m_{\left[0,t\right)};\pi,c;b^*,m^*}\vert\mathcal{F}_t\right]\leq R_t^{\xi,t,m_{\left[0,t\right)};\pi,c;b^*,m^*};
\end{equation*}
that is, for any $\left(\pi,c\right)_{\left[t,T\right)}\in\mathcal{A}_{\left[t,T\right)}$,
\begin{equation}
\begin{aligned}
&\;U\left(\xi,t;m_{\left[0,t\right)}\right)\\\geq&\; \mathbb{E}\left[U\left(X_T^{\xi,t;\pi,c;b^*,m^*},T;m^*_{\left[0,T\right)}\right)+\int_{t}^{T}U^c(c_sX_s^{\xi,t;\pi,c;b^*,m^*},s)ds\vert\mathcal{F}_t\right].
\end{aligned}
\label{new_eq_1}
\end{equation}
In turn, \eqref{new_eq_1} implies that
\begin{equation}
\begin{aligned}
&\;U\left(\xi,t;m_{\left[0,t\right)}\right)\\\geq&\; \esssup_{\left(\pi,c\right)\in\mathcal{A}}\mathbb{E}\left[U\left(X_T^{\xi,t;\pi,c;b^*,m^*},T;m^*_{\left[0,T\right)}\right)+\int_{t}^{T}U^c(c_sX_s^{\xi,t;\pi,c;b^*,m^*},s)ds\vert\mathcal{F}_t\right]\\\geq&\;\essinf_{\left(b,m\right)_{\left[t,T\right)}\in\left(\mathcal{B}\times\mathcal{P}\left(U_{\sigma}\right)\right)_{\left[t,T\right)}}\esssup_{\left(\pi,c\right)\in\mathcal{A}}\;\\&\;\mathbb{E}\left[U\left(X_T^{\xi,t;\pi,c;b,m},T;m_{\left[0,T\right)}\right)+\int_{t}^{T}U^c\left(c_sX_s^{\xi,t;\pi,c;b,m},s\right)ds\vert\mathcal{F}_t\right].
\end{aligned}
\label{new_eq_2}
\end{equation}

On the other hand, for any $\left(b,m\right)_{\left[t,T\right)}\in\left(\mathcal{B}\times\mathcal{P}\left(U_{\sigma}\right)\right)_{\left[t,T\right)}$ and $s\in\left[t,T\right)$, by Lemma \ref{lemma:H_saddle_point},
\begin{align*}
&\;H\left(s,Z^{m}_s,\bar{Z}^{m}_s;\pi^*_s;b_s,m_s\right)-H^*\left(s,Z^{m}_s,\bar{Z}^{m}_s\right)\\&+\left(c^*_s\right)^{\kappa}e^{-Y^{m}_s+g^{m}_s}\lambda_s-\kappa c^*_s-\left(1-\kappa\right)\lambda_s^{\frac{1}{1-\kappa}}e^{-\frac{Y^{m}_s}{1-\kappa}}e^{\frac{g^{m}_s}{1-\kappa}}\\=&\;H\left(s,Z^{m}_s,\bar{Z}^{m}_s;\pi^*_s;b_s,m_s\right)-H^*\left(s,Z^{m}_s,\bar{Z}^{m}_s\right)\geq 0,
\end{align*}
where $\left(\pi^*,c^*\right)_{\left[t,T\right)}\in\mathcal{A}_{\left[t,T\right)}$ is given in \eqref{eq:pi_star}, in which, for any $s\in\left[t,T\right)$, $\left(\pi^*_s,c^*_s\right)$ depends on $m_{\left[0,s\right)}=m_{\left[0,t\right)}\oplus m_{\left[t,s\right)}$. Therefore, for any $\left(b,m\right)_{\left[t,T\right)}\in\left(\mathcal{B}\times\mathcal{P}\left(U_{\sigma}\right)\right)_{\left[t,T\right)}$, $R^{\xi,t,m_{\left[0,t\right)};\pi^*,c^*;b,m}$ is a local $\mathbb{F}$-submartingale, and thus there exists an increasing sequence of $\mathbb{F}$-stopping times $\tau_n\in\left[t,T\right]$ such that $\tau_n\uparrow T$ and, in particular,
\begin{equation*}
\mathbb{E}\left[R_{\tau_n}^{\xi,t,m_{\left[0,t\right)};\pi^*,c^*;b,m}\vert\mathcal{F}_t\right]\geq R_{t}^{\xi,t,m_{\left[0,t\right)};\pi^*,c^*;b,m};
\end{equation*}
that is, for any $\left(b,m\right)_{\left[t,T\right)}\in\left(\mathcal{B}\times\mathcal{P}\left(U_{\sigma}\right)\right)_{\left[t,T\right)}$,
\begin{align*}
&\;U\left(\xi,t;m_{\left[0,t\right)}\right)\\\leq&\; \mathbb{E}\left[U\left(X_{\tau_n}^{\xi,t;\pi^*,c^*;b,m},\tau_n;m_{\left[0,\tau_n\right)}\right)+\int_{t}^{\tau_n}U^c(c^*_sX_s^{\xi,t;\pi^*,c^*;b,m},s)ds\vert\mathcal{F}_t\right],
\end{align*}
with the increasing sequence of $\mathbb{F}$-stopping times $\tau_n\in\left[t,T\right]$ such that $\tau_n\uparrow T$. Suppose, {\it at the moment}, that the class
\begin{equation*}
\left\{U\left(X_{\tau}^{\xi,t;\pi^*,c^*;b,m},\tau;m_{\left[0,\tau\right)}\right)\right\}_{\tau\in\mathcal{T}\left[t,T\right]},
\end{equation*}
where $\mathcal{T}\left[t,T\right]$ is the set of all $\mathbb{F}$-stopping times with $\tau\in\left[t,T\right]$, is uniformly integrable. By the Bounded Convergence Theorem and the Monotone Convergence Theorem, together with the fact that $U^c$ is non-negative, for any $\left(b,m\right)_{\left[t,T\right)}\in\left(\mathcal{B}\times\mathcal{P}\left(U_{\sigma}\right)\right)_{\left[t,T\right)}$,
\begin{align*}
&\;U\left(\xi,t;m_{\left[0,t\right)}\right)\\\leq&\; \lim_{n\uparrow\infty}\mathbb{E}\left[U\left(X_{\tau_n}^{\xi,t;\pi^*,c^*;b,m},\tau_n;m_{\left[0,\tau_n\right)}\right)+\int_{t}^{\tau_n}U^c(c^*_sX_s^{\xi,t;\pi^*,c^*;b,m},s)ds\vert\mathcal{F}_t\right]\\=&\;\mathbb{E}\left[U\left(X_{T}^{\xi,t;\pi^*,c^*;b,m},T;m_{\left[0,T\right)}\right)+\int_{t}^{T}U^c(c^*_sX_s^{\xi,t;\pi^*,c^*;b,m},s)ds\vert\mathcal{F}_t\right];
\end{align*}
that is, $R^{\xi,t,m_{\left[0,t\right)};\pi^*,c^*;b,m}$ is even a proper $\mathbb{F}$-submartingale. Therefore, for any $\left(b,m\right)_{\left[t,T\right)}\in\left(\mathcal{B}\times\mathcal{P}\left(U_{\sigma}\right)\right)_{\left[t,T\right)}$,
\begin{equation}
\begin{aligned}
&\;U\left(\xi,t;m_{\left[0,t\right)}\right)\\\leq&\;\mathbb{E}\left[U\left(X_{T}^{\xi,t;\pi^*,c^*;b,m},T;m_{\left[0,T\right)}\right)+\int_{t}^{T}U^c(c^*_sX_s^{\xi,t;\pi^*,c^*;b,m},s)ds\vert\mathcal{F}_t\right].
\label{new_eq_3}
\end{aligned}
\end{equation}
In turn, \eqref{new_eq_3} implies that
\begin{equation}
\begin{aligned}
&\;U\left(\xi,t;m_{\left[0,t\right)}\right)\\\leq&\;\essinf_{\left(b,m\right)_{\left[t,T\right)}\in\left(\mathcal{B}\times\mathcal{P}\left(U_{\sigma}\right)\right)_{\left[t,T\right)}}\\&\;\mathbb{E}\left[U\left(X_{T}^{\xi,t;\pi^*,c^*;b,m},T;m_{\left[0,T\right)}\right)+\int_{t}^{T}U^c(c^*_sX_s^{\xi,t;\pi^*,c^*;b,m},s)ds\vert\mathcal{F}_t\right]\\\leq&\;\esssup_{\left(\pi,c\right)\in\mathcal{A}}\;\essinf_{\left(b,m\right)_{\left[t,T\right)}\in\left(\mathcal{B}\times\mathcal{P}\left(U_{\sigma}\right)\right)_{\left[t,T\right)}}\\&\;\mathbb{E}\left[U\left(X_T^{\xi,t;\pi,c;b,m},T;m_{\left[0,T\right)}\right)+\int_{t}^{T}U^c\left(c_sX_s^{\xi,t;\pi,c;b,m},s\right)ds\vert\mathcal{F}_t\right].
\label{new_eq_4}
\end{aligned}
\end{equation}

Therefore, \eqref{new_eq_1}, \eqref{new_eq_2}, \eqref{new_eq_3}, and \eqref{new_eq_4} together yield the results. Finally, it remains to show that the class
\begin{equation*}
\left\{\frac{\left(X_{\tau}^{\xi,t;\pi^*,c^*;b,m}\right)^{\kappa}}{\kappa}e^{Y^m_\tau-g^m_\tau}\right\}_{\tau\in\mathcal{T}\left[t,T\right]}
\end{equation*}
is uniformly integrable. By Proposition \ref{prop:power_infinite_BSDE_existence_drift_vol}, $Y^m$ is uniformly bounded. Moreover, $g^m$ is also uniformly bounded in $\left[t,T\right]$. Therefore, it suffices to show that the class $\left\{\left(X_{\tau}^{\xi,t;\pi^*,c^*;b,m}\right)^{\kappa}\right\}_{\tau\in\mathcal{T}\left[t,T\right]}$ is uniformly integrable. Since $X^{\xi,t;\pi^*,c^*;b,m}$ solves \eqref{wealth_random_2}, for any $s\geq t$ and $\left(b,m\right)_{\left[t,s\right)}\in\left(\mathcal{B}\times\mathcal{P}\left(U_{\sigma}\right)\right)_{\left[t,s\right)}$,
\begin{align*}
\left(X_{s}^{\xi,t;\pi^*,c^*;b,m}\right)^{\kappa}=&\;\xi^{\kappa}\times\exp\Bigg(\int_{t}^{s}\kappa\left(r+\sum_{i=1}^{n}\pi_v^{i,*}\left(b_v^i-r\right)-c^*_v\right)\\&\;\quad\quad+\frac{1}{2}\kappa\left(\kappa-1\right)\sum_{i=1}^{n}\left(\pi_v^{i,*}\right)^2\left(\int_{U_{\sigma}^i}u^2m_v^i\left(du\right)+\left(\bar{\sigma}_v^i\right)^2\right)dv\Bigg)\\&\;\quad\times\mathcal{E}\left(\kappa\sum_{i=1}^{n}\int_{t}^{\cdot}\pi_v^{i,*}\left(\sqrt{\int_{U_{\sigma}^i}u^2m_v^i\left(du\right)}dW_v^i+\bar{\sigma}_v^id\bar{W}_v\right)\right)_{s},
\end{align*}
where $\mathcal{E}\left(\cdot\right)_s$, $s\geq t$, is the Dol{\'e}ans-Dade exponential. By the compactness of $\Pi$ and $U_{\sigma}$, as well as the uniform boundedness of $\bar{\sigma}$, the Dol{\'e}ans-Dade exponential is a uniformly integrable martingale. Since {$Y^m$, $g^m$, and $\lambda$} are uniformly bounded in $\left[t,T\right]$, $c^*$ is also uniformly bounded. Hence, together with the compactness of $U_b$, the class $\left\{\left(X_{\tau}^{\xi,t;\pi^*,c^*;b,m}\right)^{\kappa}\right\}_{\tau\in\mathcal{T}\left[t,T\right]}$ is indeed uniformly integrable.\\

{\subsection{Verification for Robust Forward Preference: Proof of Proposition \ref{new_prop}}}
Fix $t\geq 0$, $\xi\in\mathcal{L}\left(\mathcal{F}_t;\mathbb{R}_+\right)$, $m_{\left[0,t\right)}\in\mathcal{P}\left(U_{\sigma}\right)_{\left[0,t\right)}$, and $T\geq t$, throughout this proof.

Notice that, for any $\left(\omega,s,z,\bar{z}\right)\in\Omega\times[0,\infty)\times\mathbb{R}^n\times\mathbb{R}$ and $x_{\pi}\in\Pi$,
\begin{equation}
\begin{aligned}
&\;\inf_{\left(x_{b},x_{m}\right)\in U_b\times\mathcal{P}\left(U_{\sigma}\right)_{\omega,s}}H\left(\omega,s,z,\bar{z};x_{\pi};x_{b},x_{m}\right)\\=&\;\inf_{\left(x_{b},x_{m}\right)\in U_b\times\mathcal{P}'\left(U_{\sigma}\right)_{\omega,s}}H\left(\omega,s,z,\bar{z};x_{\pi};x_{b},x_{m}\right)\\=&\;\inf_{\left(x_{b},x_{\sigma}\right)\in U_b\times U_{\sigma}}H\left(\omega,s,z,\bar{z};x_{\pi};x_{b},x_{\sigma}\right),
\end{aligned}
\label{eq:H_equivalence}
\end{equation}
where the functions $H$ are defined in \eqref{eq:H} and \eqref{eq:H_canonical}. The second equality is true by definition. For the first equality, since $\mathcal{P}'\left(U_{\sigma}\right)_{\omega,s}\subseteq\mathcal{P}\left(U_{\sigma}\right)_{\omega,s}$,
\begin{align*}
&\;\inf_{\left(x_{b},x_{m}\right)\in U_b\times\mathcal{P}\left(U_{\sigma}\right)_{\omega,s}}H\left(\omega,s,z,\bar{z};x_{\pi};x_{b},x_{m}\right)\\\leq&\;\inf_{\left(x_{b},x_{m}\right)\in U_b\times\mathcal{P}'\left(U_{\sigma}\right)_{\omega,s}}H\left(\omega,s,z,\bar{z};x_{\pi};x_{b},x_{m}\right);
\end{align*}
moreover, for any $\left(x_{b},x_{m}\right)\in U_b\times\mathcal{P}\left(U_{\sigma}\right)_{\omega,s}$,
\begin{align*}
H\left(\omega,s,z,\bar{z};x_{\pi};x_{b},x_{m}\right)=&\;\int_{U_{\sigma}}H\left(\omega,s,z,\bar{z};x_{\pi};x_{b},u\right)x_m\left(du\right)\\\geq&\;\int_{U_{\sigma}}\inf_{\left(x_{b},x_{\sigma}\right)\in U_b\times U_{\sigma}}H\left(\omega,s,z,\bar{z};x_{\pi};x_{b},x_{\sigma}\right)x_m\left(du\right)\\=&\;\inf_{\left(x_{b},x_{\sigma}\right)\in U_b\times U_{\sigma}}H\left(\omega,s,z,\bar{z};x_{\pi};x_{b},x_{\sigma}\right)\\=&\;\inf_{\left(x_{b},x_{m}\right)\in U_b\times\mathcal{P}'\left(U_{\sigma}\right)_{\omega,s}}H\left(\omega,s,z,\bar{z};x_{\pi};x_{b},x_{m}\right),
\end{align*}
and hence
\begin{align*}
&\;\inf_{\left(x_{b},x_{m}\right)\in U_b\times\mathcal{P}\left(U_{\sigma}\right)_{\omega,s}}H\left(\omega,s,z,\bar{z};x_{\pi};x_{b},x_{m}\right)\\\geq&\;\inf_{\left(x_{b},x_{m}\right)\in U_b\times\mathcal{P}'\left(U_{\sigma}\right)_{\omega,s}}H\left(\omega,s,z,\bar{z};x_{\pi};x_{b},x_{m}\right).
\end{align*}
Therefore, Lemma \ref{lemma:H_saddle_point} and \eqref{eq:H_equivalence} together imply that, for any $\left(\pi,c\right)_{\left[t,T\right)}\in\mathcal{A}_{\left[t,T\right)}$ and $s\in\left[t,T\right)$,
\begin{align*}
0\geq&\;\inf_{\left(x_{b},x_{m}\right)\in U_b\times\mathcal{P}\left(U_{\sigma}\right)_{\omega,s}}H\left(s,Z^{\delta^*}_s,\bar{Z}^{\delta^*}_s;\pi_s;x_{b},x_{m}\right)-H^*\left(s,Z^{\delta^*}_s,\bar{Z}^{\delta^*}_s\right)\\=&\;\inf_{\left(x_{b},x_{m}\right)\in U_b\times\mathcal{P}'\left(U_{\sigma}\right)_{\omega,s}}H\left(s,Z^{\delta^*}_s,\bar{Z}^{\delta^*}_s;\pi_s;x_{b},x_{m}\right)-H^*\left(s,Z^{\delta^*}_s,\bar{Z}^{\delta^*}_s\right)\\=&\;H\left(s,Z^{\delta^*}_s,\bar{Z}^{\delta^*}_s;\pi_s;b^*_s,\delta^*_s\right)-H^*\left(s,Z^{\delta^*}_s,\bar{Z}^{\delta^*}_s\right),
\end{align*}
and
\begin{equation*}
0\geq c_s^{\kappa}e^{-Y^{\delta^*}_s+g^{\delta^*}_s}\lambda_s-\kappa c_s-\left(1-\kappa\right)\lambda_s^{\frac{1}{1-\kappa}}e^{-\frac{Y^{\delta^*}_s}{1-\kappa}}e^{\frac{g^{\delta^*}_s}{1-\kappa}};
\end{equation*}
herein, $b^*_s=x^*_{b}\left(s,Z^{\delta^*}_s,\bar{Z}^{\delta^*}_s\right)$ and $\delta^*_s=x^*_{\delta}\left(s,Z^{\delta^*}_s,\bar{Z}^{\delta^*}_s\right)$, where $\left(x^*_{b},x^*_{\delta}\right)\in U_b\times\mathcal{P}'\left(U_{\sigma}\right)_{\omega,s}$ exists due to the compactness of $U_b$ and $U_{\sigma}$ (with $x^*_{\delta}$ denoting $x^*_{m}\in\mathcal{P}'\left(U_{\sigma}\right)_{\omega,s}$ to recognize that the Borel probability measure is in fact a Dirac measure), $Z^{\delta^*}_s,\bar{Z}^{\delta^*}_s,Y^{\delta^*}_s,g^{\delta^*}_s$ depend on $\delta^*_{\left[0,s\right)}=m_{\left[0,t\right)}\oplus\delta^*_{\left[t,s\right)}$, and the saddle value function $H^*$ is defined in \eqref{eq:saddle_valeu_equivalence}. 
Together with \eqref{eq:R}, since $R^{\xi,t,m_{\left[0,t\right)};\pi,c;b^*,\delta^*}$ is non-negative, for any $\left(\pi,c\right)_{\left[t,T\right)}\in\mathcal{A}_{\left[t,T\right)}$, $R^{\xi,t,m_{\left[0,t\right)};\pi,c;b^*,\delta^*}$ is a proper $\mathbb{F}$-supermartingale. Hence, for any $\left(\pi,c\right)_{\left[t,T\right)}\in\mathcal{A}_{\left[t,T\right)}$,
\begin{equation*}
\mathbb{E}\left[R_T^{\xi,t,m_{\left[0,t\right)};\pi,c;b^*,\delta^*}\vert\mathcal{F}_t\right]\leq R_t^{\xi,t,m_{\left[0,t\right)};\pi,c;b^*,\delta^*};
\end{equation*}
that is, for any $\left(\pi,c\right)\in\mathcal{A}$,
\begin{equation}
\begin{aligned}
&\;U\left(\xi,t;m_{\left[0,t\right)}\right)\\\geq&\;\essinf_{\left(b,m\right)_{\left[t,T\right)}\in\left(\mathcal{B}\times\mathcal{P}'\left(U_{\sigma}\right)\right)_{\left[t,T\right)}}\\&\;\mathbb{E}\left[U\left(X_{T}^{\xi,t;\pi,c;b,m},T;m_{\left[0,T\right)}\right)+\int_{t}^{T}U^c(c_sX_s^{\xi,t;\pi,c;b,m},s)ds\vert\mathcal{F}_t\right],
\end{aligned}
\label{eq:withoutsup}
\end{equation}
which implies that
\begin{equation}
\begin{aligned}
&\;U\left(\xi,t;m_{\left[0,t\right)}\right)\\\geq&\;\esssup_{\left(\pi,c\right)\in\mathcal{A}}\essinf_{\left(b,m\right)_{\left[t,T\right)}\in\left(\mathcal{B}\times\mathcal{P}'\left(U_{\sigma}\right)\right)_{\left[t,T\right)}}\\&\;\mathbb{E}\left[U\left(X_{T}^{\xi,t;\pi,c;b,m},T;m_{\left[0,T\right)}\right)+\int_{t}^{T}U^c(c_sX_s^{\xi,t;\pi,c;b,m},s)ds\vert\mathcal{F}_t\right].
\end{aligned}
\label{eq:supposed_relation}
\end{equation}
Note that \eqref{eq:withoutsup} particularly holds true with $\left(\pi^*,c^*\right)\in\mathcal{A}$, which is given by \eqref{eq:pi_star}.

To prove the first equality in \eqref{eq:summary_expand}, on one hand, \eqref{eq:withoutsup} and \eqref{eq:summary} together imply that
\begin{align*}
&\;\essinf_{\left(b,m\right)_{\left[t,T\right)}\in\left(\mathcal{B}\times\mathcal{P}\left(U_{\sigma}\right)\right)_{\left[t,T\right)}}\\&\;\mathbb{E}\left[U\left(X_{T}^{\xi,t;\pi^*,c^*;b,m},T;m_{\left[0,T\right)}\right)+\int_{t}^{T}U^c(c^*_sX_s^{\xi,t;\pi^*,c^*;b,m},s)ds\vert\mathcal{F}_t\right]\\\geq&\;\essinf_{\left(b,m\right)_{\left[t,T\right)}\in\left(\mathcal{B}\times\mathcal{P}'\left(U_{\sigma}\right)\right)_{\left[t,T\right)}}\\&\;\mathbb{E}\left[U\left(X_{T}^{\xi,t;\pi^*,c^*;b,m},T;m_{\left[0,T\right)}\right)+\int_{t}^{T}U^c(c^*_sX_s^{\xi,t;\pi^*,c^*;b,m},s)ds\vert\mathcal{F}_t\right],
\end{align*}
On the other hand, since $\mathcal{P}'\left(U_{\sigma}\right)_{\left[t,T\right)}\subseteq\mathcal{P}\left(U_{\sigma}\right)_{\left[t,T\right)}$,
\begin{align*}
&\;\essinf_{\left(b,m\right)_{\left[t,T\right)}\in\left(\mathcal{B}\times\mathcal{P}\left(U_{\sigma}\right)\right)_{\left[t,T\right)}}\\&\;\mathbb{E}\left[U\left(X_{T}^{\xi,t;\pi^*,c^*;b,m},T;m_{\left[0,T\right)}\right)+\int_{t}^{T}U^c(c^*_sX_s^{\xi,t;\pi^*,c^*;b,m},s)ds\vert\mathcal{F}_t\right]\\\leq&\;\essinf_{\left(b,m\right)_{\left[t,T\right)}\in\left(\mathcal{B}\times\mathcal{P}'\left(U_{\sigma}\right)\right)_{\left[t,T\right)}}\\&\;\mathbb{E}\left[U\left(X_{T}^{\xi,t;\pi^*,c^*;b,m},T;m_{\left[0,T\right)}\right)+\int_{t}^{T}U^c(c^*_sX_s^{\xi,t;\pi^*,c^*;b,m},s)ds\vert\mathcal{F}_t\right].
\end{align*}
In both inequalities, $\left(\pi_s^*,c_s^*\right)$, for $s\in\left[t,T\right)$, on the left hand side depends on $m_{\left[0,s\right)}=m_{\left[0,t\right)}\oplus m_{\left[t,s\right)}$ with $m_{\left[t,s\right)}\in\mathcal{P}\left(U_{\sigma}\right)_{\left[t,s\right)}$, while that on the right hand side depends on $m_{\left[0,s\right)}=m_{\left[0,t\right)}\oplus m_{\left[t,s\right)}$ with $m_{\left[t,s\right)}\in\mathcal{P}'\left(U_{\sigma}\right)_{\left[t,s\right)}$. These imply the first equality in \eqref{eq:summary_expand}.

To prove the second equality in \eqref{eq:summary_expand}, on one hand, by \eqref{eq:summary} and since $\mathcal{P}'\left(U_{\sigma}\right)_{\left[t,T\right)}\subseteq\mathcal{P}\left(U_{\sigma}\right)_{\left[t,T\right)}$,
\begin{align*}
&\;\essinf_{\left(b,m\right)_{\left[t,T\right)}\in\left(\mathcal{B}\times\mathcal{P}\left(U_{\sigma}\right)\right)_{\left[t,T\right)}}\;\esssup_{\left(\pi,c\right)\in\mathcal{A}}\\&\;\mathbb{E}\left[U\left(X_T^{\xi,t;\pi,c;b,m},T;m_{\left[0,T\right)}\right)+\int_{t}^{T}U^c\left(c_sX_s^{\xi,t;\pi,c;b,m},s\right)ds\vert\mathcal{F}_t\right]\\=&\;\essinf_{\left(b,m\right)_{\left[t,T\right)}\in\left(\mathcal{B}\times\mathcal{P}\left(U_{\sigma}\right)\right)_{\left[t,T\right)}}\\&\;\mathbb{E}\left[U\left(X_{T}^{\xi,t;\pi^*,c^*;b,m},T;m_{\left[0,T\right)}\right)+\int_{t}^{T}U^c(c^*_sX_s^{\xi,t;\pi^*,c^*;b,m},s)ds\vert\mathcal{F}_t\right]\\\leq&\;\essinf_{\left(b,m\right)_{\left[t,T\right)}\in\left(\mathcal{B}\times\mathcal{P}'\left(U_{\sigma}\right)\right)_{\left[t,T\right)}}\\&\;\mathbb{E}\left[U\left(X_{T}^{\xi,t;\pi^*,c^*;b,m},T;m_{\left[0,T\right)}\right)+\int_{t}^{T}U^c(c^*_sX_s^{\xi,t;\pi^*,c^*;b,m},s)ds\vert\mathcal{F}_t\right]\\\leq&\;\esssup_{\left(\pi,c\right)\in\mathcal{A}}\;\essinf_{\left(b,m\right)_{\left[t,T\right)}\in\left(\mathcal{B}\times\mathcal{P}'\left(U_{\sigma}\right)\right)_{\left[t,T\right)}}\\&\;\mathbb{E}\left[U\left(X_T^{\xi,t;\pi,c;b,m},T;m_{\left[0,T\right)}\right)+\int_{t}^{T}U^c\left(c_sX_s^{\xi,t;\pi,c;b,m},s\right)ds\vert\mathcal{F}_t\right].
\end{align*}
On the other hand, by \eqref{eq:supposed_relation} and \eqref{eq:summary},
\begin{align*}
&\;\esssup_{\left(\pi,c\right)\in\mathcal{A}}\;\essinf_{\left(b,m\right)_{\left[t,T\right)}\in\left(\mathcal{B}\times\mathcal{P}'\left(U_{\sigma}\right)\right)_{\left[t,T\right)}}\\&\;\mathbb{E}\left[U\left(X_T^{\xi,t;\pi,c;b,m},T;m_{\left[0,T\right)}\right)+\int_{t}^{T}U^c\left(c_sX_s^{\xi,t;\pi,c;b,m},s\right)ds\vert\mathcal{F}_t\right]\\\leq&\;\essinf_{\left(b,m\right)_{\left[t,T\right)}\in\left(\mathcal{B}\times\mathcal{P}\left(U_{\sigma}\right)\right)_{\left[t,T\right)}}\;\esssup_{\left(\pi,c\right)\in\mathcal{A}}\\&\;\mathbb{E}\left[U\left(X_T^{\xi,t;\pi,c;b,m},T;m_{\left[0,T\right)}\right)+\int_{t}^{T}U^c\left(c_sX_s^{\xi,t;\pi,c;b,m},s\right)ds\vert\mathcal{F}_t\right].
\end{align*}
These imply the second equality in \eqref{eq:summary_expand}.\\

\section{Conclusions}

This paper investigates robust forward investment and consumption preferences under drift and volatility uncertainties, focusing on optimal and robust strategies for a risk-averse and ambiguity-averse agent in an incomplete financial market. We construct non-zero volatility robust CRRA forward preferences and associated optimal strategies in a physical market where drifts and idiosyncratic volatilities are uncertain, but the systematic volatility is known and stochastic. The Hamiltonian of the primal maximin problem lacks convexity with respect to idiosyncratic volatility, generally precluding a saddle point and rendering standard robust preference construction via saddle points infeasible.

To address this, we employ randomization, endogenously randomizing the idiosyncratic volatility process to define an auxiliary financial market. By also randomizing the unhedgeable risks, we ensure the randomized Hamiltonian admits a saddle point. The saddle value is used to construct a randomized infinite-horizon BSDE and an ODE, whose unique solutions yield the non-zero volatility robust randomized CRRA forward preferences and optimal strategies in the auxiliary market. We prove that these preferences and strategies are also robust and optimal in the physical market, adapting dynamically to the realized idiosyncratic volatility. At each future time, the agent implements an optimal strategy derived from the saddle point of the randomized Hamiltonian, accounting for worst-case drift and randomization.

The randomization approach via randomized BSDEs is of independent interest. It would be interesting to compare this method to the 2BSDE approach for optimal investment under static preferences in \cite{Matoussi_2015} and the G-expectation BSDE approach for forward preferences in \cite{Sun_2024}.

\end{document}